\documentclass[a4paper,10pt]{article}
\usepackage{amsmath}
\usepackage{parskip}
\usepackage{ucs}

\usepackage[utf8]{inputenc} 
\usepackage[pdftex]{graphicx}

\usepackage{subcaption}
\graphicspath{{./Pic/}}
\DeclareGraphicsExtensions{.pdf,.png,.jpg}
\usepackage{ulem}
\usepackage{hhline}
\usepackage{color,colortbl}
\usepackage{authblk}

\title{
Equations and improved coefficients for parallel transport in multicomponent collisional plasmas: method and application for tokamak modelling}

\author[1]{S. Makarov}
\author[1]{D. Coster}
\author[2]{V. Rozhansky}
\author[3]{A. Stepanenko}
\author[2]{E. Kaveeva}
\author[2]{I. Senichenkov}
\author[3]{V. Zhdanov}
\author[4]{X. Bonnin}

\affil[1]{Max-Planck-Institut f\"ur Plasmaphysik, D-85748 Garching, Germany}
\affil[2]{Peter the Great St.Petersburg Polytechnic University, St.Petersburg, Russia}
\affil[3]{National Research Nuclear University MEPhI (Moscow Engineering Physics Institute), Kashirskoe sh. 31,
115409, Moscow, Russia}
\affil[4]{ITER Organization, CS 90 046, F-13067, St-Paul-Lez-Durance Cedex, France}

\begin{document}

\maketitle

\begin{abstract}
New analytical expressions for parallel transport coefficients in multicomponent collisional plasmas are presented in this paper. They are improved versions of the expressions written in [V. M. Zhdanov. Transport Processes in Multicomponent Plasma, vol. 44. 10 2002.]\nocite{book}, based on Grad's 21N-moment method. Both explicit and approximate approaches for transport coefficients calculation are considered.  Accurate application of this closure for the Braginskii transport equations is discussed. Viscosity dependence on the heat flux is taken into account. Improved expressions are implemented into the SOLPS-ITER code and tested for deuterium and neon ITER cases. Some typos found in [V. M. Zhdanov. Transport Processes in Multicomponent Plasma, vol. 44. 10 2002.]\nocite{book} are corrected.
\end{abstract}
\section{Introduction}
Fusion toroidal devices with magnetic confinement (tokamaks and stellarators) operate with multispecies plasma. Plasma composition can contain mixtures of: main components: hydrogen, deuterium, tritium \cite{Watkins__1999} and helium isotopes ; and impurities: helium, lithium, beryllium, carbon, nitrogen, neon, argon, etc.\  \cite{Kallenbach_2020}. (In some operational regimes, helium might be the main component and then, in addition to the usual impurities, isotopes of hydrogen would be impurities.) 
Edge plasmas usually are in a collisional regime, by which we mean that macroscopic parameters change slowly on parallel (perpendicular) length scales of the mean free path (gyroradius) and time scales between collisions (gyromotion period).
Under this assumption,
a closure method such as the one proposed by Braginskii \cite{braginskii1965transport} can be applied to the moments of the distribution function. However, this approach is applied for a single ion species case and assumes only trace levels of impurities. For the multicomponent case, Grad's method \cite{grad1963asymptotic} can be used. This method is based on the tensorial Hermite polynomials finite expansion approximation of the distribution function with a local Maxwellian distribution function as the zeroth-order approximation. It is developed in detail in \cite{book}. Based on it, the  21N-moment approximation is studied in this work, though only parallel components are considered in this paper. Equations for electrons can be solved separately due to the large electron-ion mass difference, as it was considered in \cite{book}. For ions the situation is more complicated, especially for mixtures with comparable masses. This is analyzed in detail below.     

Some previous work on implementing the moment approach can be found in \cite{bergmann1996implementation} (based on \cite{igitkhanov1988impurity,igitkhanov1994implication} and described in \cite{schneider2006plasma}) where an explicit implementation (based on matrix inversions) was implemented in SOLPS;  \cite{BUFFERAND201982} where the multispecies closure was implemented into 3D turbulence and transport codes based on \cite{book}; and additional work in \cite{fomin2017transport}.
Some of this work assumed $m_{light}/m_{heavy}\ll1$ and so could not be applied to deuterium, tritium and helium mixture cases, while in others the closure for viscosity was not considered.  In addition, there is a difference between definitions made by Braginskii \cite{braginskii1965transport} and in \cite{book}. Thus, for the application of the closures discussed in \cite{book} for the Braginskii equations, special corrections are developed in this paper (see Section \ref{section:fluid_eq}).  Some typos in \cite{book} which directly affect the results are corrected in appendix \ref{app:3}. 
In \cite{bergmann1996implementation}, a full closure (including viscosity) was implemented into the transport code, though the heat flux dependent part of viscosity was not considered and this higher order effect plays an important role in toroidal systems \cite{helander2005collisional,Hirshman_1981}. Such viscosity was added by Rozhansky et al \cite{Rozhansky_2001, Rozhansky_2009} for correct calculation of the radial electric field in the H-mode pedestal in tokamaks, but only in the single fluid case and can be extended to multispecies cases.

In this paper this generalization of the equation system is presented. Section \ref{section:fluid_eq} is dedicated to an accurate application of Grad's closure discussed in \cite{book} for the Braginskii system of equations used in transport codes, for instance SOLPS-ITER \cite{WIESEN2015480,Bonnin2016}. In addition, the heat flux dependent part of the viscosity is considered in section \ref{section:fluid_eq}.

Since the calculation of transport coefficients is supposed to be performed during the numerical solution of algebraic equations, this might be a problem in a fluid code there this would be done at each time-step in each grid cell; therefore new improved analytical expressions are developed in  section \ref{section:Analyt}. In the case where masses of species are significantly different, such methods can be applied. This method is an improvement of the method discussed in \cite{book}. Comparison for the improved expressions, the original expressions (8.4.7) in \cite{book}, and the solution of the explicit system of algebraic equations is provided.
As it is shown there, even for the cases, where masses of components are not too different, analytical formulae provide surprisingly good agreement to the explicit numerical approach. Moreover, for some cases, the new formulae provide a much better match with the numerical results than the original analytical expressions (8.4.7) in \cite{book}. Results can be found in appendix \ref{app:1}.

Then, in section \ref{For_SOlPS}, a comparison between transport coefficients calculated using Grad's 21N-moment closure and transport coefficients used in the current SOLPS-ITER model is considered.  Additionally, new expressions for thermal and friction forces were implemented into the SOLPS-ITER code and tested for neon transport in an ITER deuterium plasma and compared with the current SOLPS-ITER model \cite{Sytova2018,Sytova2020}.
\section{Basic equations for multispecies plasma}\label{section:fluid_eq}
\subsection{Equations}

First of all, let's consider the system of fluid equations for species type $\alpha$ and charge state $Z$, which can be applied for both ions and electrons (for electrons $Z=-1$):
\begin{align}\label{eq:density_eq}
\partial_tn_{\alpha Z}+\nabla\cdot(\textbf u_{\alpha Z}n_{\alpha Z})=S^n_{\alpha Z}
\end{align}
\begin{multline}\label{eq:momentum_eq}
m_{\alpha}\partial_t(\textbf u_{\alpha Z}n_{\alpha Z})+{\nabla}\cdot\overleftrightarrow{\Gamma}^m_{\alpha Z}=\\-\nabla (n_{\alpha Z}T^{(Br)}_{\alpha Z})+Zen_{\alpha Z}(\textbf E+[\textbf u_{\alpha Z}\textbf B])-\nabla \cdot \overleftrightarrow{\pi}^{(Br)}_{\alpha Z}+\textbf R_{\alpha Z}+
\textbf S^{m}_{\alpha Z}
\end{multline}
\begin{multline}\label{eq:ion_heat_eq}
\frac{3}{2}\partial_t(n_{\alpha Z}T^{(Br)}_{\alpha Z})+\nabla\cdot(
\textbf q^{(Br)}_{\alpha Z}+\frac{3}{2}\textbf u_{\alpha Z}n_{\alpha Z} T^{(Br)}_{\alpha Z})+{n_{\alpha Z} T^{(Br)}_{\alpha Z}}\nabla \cdot\textbf u_{\alpha Z}=\\
-((\overleftrightarrow{\pi}^{(Br)}_{\alpha Z}\nabla)\textbf u_{\alpha Z}) +Q^{(Br)}_{\alpha Z}+
S^{E}_{\alpha Z}
\end{multline}
Details can be found in \cite{braginskii1965transport}. In this equation system temperatures, heat fluxes and viscosities are defined according to Braginskii:
\begin{align}\label{eq:def_Br_T1}
T^{(Br)}_{\alpha Z}=\frac{2}{3}\frac{m_\alpha}{n_{\alpha Z}} \iiint\frac{(\textbf v-\textbf u_{\alpha Z})^2}{2}f_{\alpha Z }d^3\textbf v
\end{align}
\begin{align}\label{eq:def_Br_heat1}
q^{(Br)}_{\alpha Z k}=m_\alpha \iiint\frac{(\textbf v-\textbf u_{\alpha Z})^2}{2}(v_k-u_{\alpha Z k})f_{\alpha Z }d^3\textbf {v}
\end{align}
\begin{align}\label{eq:def_Br_vis1}
\pi^{(Br)}_{\alpha Zkl}=m_\alpha \iiint \Big[(v_k-u_{\alpha Z k})(v_l-u_{\alpha Zl})-\frac{\delta_{kl}}{3}(\textbf v-\textbf u_{\alpha Z})^2\Big]f_{\alpha Z }d^3\textbf {v}
\end{align}
where $f_{\alpha Z }$ is the distribution function for ion or electron species ${\alpha Z }$, k and l are component indices, and integrals of the collisional term $C_{\alpha Z }$ due to Coulomb (elastic) collisions provide:
\begin{align}\label{eq:def_R}
R_{\alpha Zk}=m_\alpha \iiint v_kC_{\alpha Z }d^3\textbf {v}
\end{align}
\begin{align}\label{eq:def_Q}
Q^{(Br)}_{\alpha Z}=m_\alpha \iiint \frac{(\textbf v-\textbf u_{\alpha Z})^2}{2}C_{\alpha Z }d^3\textbf {v}
\end{align}
Terms $S^n_{\alpha Z},\ S^{m}_{\alpha Zi}, S^{E}_{\alpha Z}$ describe, respectively, particle, momentum, and energy sources due to 
inelastic collisions between charged species (ionization, recombination, excitation) and all
interactions with neutrals, and they are considered as external parameters in this model.

The momentum flux due to the flow velocity is:
\begin{align}
\Gamma^m_{\alpha Zkl}=m_{\alpha}u_{\alpha Zk}u_{\alpha Zl}n_{\alpha Z}
\end{align}
The density and flow velocity are defined in appendix \ref{app:1}.

In a multispecies plasma, the transport coefficients, which help to express
$\textbf q^{(Br)}_{\alpha Z},$  $\overleftrightarrow{\pi}^{(Br)}_{\alpha Z},\textbf R_{\alpha Zi}$ and $Q^{(Br)}_{\alpha Z}$ 
through densities, velocities, temperatures, and their spatial derivatives, differ significantly for different plasma species, different species densities and different mass ratios of species. This means that no simplification may be made in the general case, and general methods for the solution of kinetic equations should be used.

Such a method is Grad's method of 21N-moment, described in \cite{book}. Following \cite{book}, we define:
\begin{align}\label{eq:def_Zhd_temp}
T_{\alpha Z}=\frac{2}{3}\frac{m_\alpha}{n_{\alpha Z}} \iiint\frac{(\textbf v-\textbf u)^2}{2}f_{\alpha Z }d^3\textbf v
\end{align}
\begin{align}\label{eq:def_Zhd_heat}
h_{\alpha Zk}=m_\alpha \iiint\frac{(\textbf v-\textbf u)^2}{2}(v_k-u_k)f_{\alpha Z }d^3\textbf v -\frac{5}{2}w_{\alpha Zk}n_{\alpha Z}T_{\alpha Z}
\end{align}
\begin{align}\label{eq:def_Zhd_vis}
\pi_{\alpha Zkl}=m_\alpha \iiint \Big[(v_k-u_{ k})(v_l-u_{l})-\frac{\delta_{kl}}{3}(\textbf v-\textbf u)^2\Big]f_{\alpha Z }d^3\textbf {v}
\end{align}
where:
\begin{align}\label{eq:def_Zhd_vel}
\textbf w_{\alpha Z}=\textbf u_{\alpha Z}-\textbf u
;\ \ \ \ 
\textbf u=\frac{\sum_{\alpha,Z}m_{\alpha}n_{\alpha Z}\textbf u_{\alpha Z }}{\sum_{\alpha,Z}m_{\alpha}n_{\alpha Z}},
\end{align}
One can note that temperature \eqref{eq:def_Zhd_temp}, heat flux \eqref{eq:def_Zhd_heat} and viscosity \eqref{eq:def_Zhd_vis} are defined in \cite{book} regarding to the mass-averaged flow velocity \eqref{eq:def_Zhd_vel}, while \eqref{eq:def_Br_T1}-\eqref{eq:def_Br_vis1} are defined with respect to the flow velocity of corresponding species. Definitions \eqref{eq:def_Zhd_temp}-\eqref{eq:def_Zhd_vis} are more suitable for this method, therefore they were used in \cite{book}. Thus, the result of the closure discussed in \cite{book}, that is expressed in heat flux, viscosity, friction term and heat exchange term, can be applied for Braginskii system of equations \eqref{eq:density_eq}-\eqref{eq:ion_heat_eq} using corrections due to definition difference.  

Indeed:
\begin{align}\label{eq:temp_diff}
T_{\alpha Z}=T^{(Br)}_{\alpha Z}+\frac{1}{3}m_\alpha\textbf w^2_{\alpha Z}
\end{align}
\begin{align}\label{eq:hf_diff}
q^{(Br)}_{\alpha Z  k}=h_{\alpha Z  k}+n_{\alpha Z}w_{\alpha Z k}m_\alpha\textbf w^2_{\alpha Z}-\sum_s w_{\alpha Z s}\pi_{\alpha Z sk}
\end{align}
\begin{align}\label{eq:pi_diff}
\pi^{(Br)}_{\alpha Zkl}=\pi_{\alpha Zkl}-m_\alpha n_{\alpha Z} w_{\alpha Z k}w_{\alpha Zl}+m_\alpha n_{\alpha Z}\frac{\delta_{kl}}{3}\textbf w_{\alpha Z}^2
\end{align}
A similar correction should be applied to the heat exchange term:
\begin{align}\label{eq:Q_diff}
Q^{(Br)}_{\alpha Z}=Q_{\alpha Z}-\textbf w_{\alpha Z}\textbf R_{\alpha Z}
\end{align}
where according to \cite{book}:
\begin{align}\label{eq:Q_Zhd}
Q_{\alpha Z}\equiv m_\alpha \iiint \frac{(\textbf v-\textbf u)^2}{2}C_{\alpha Z }d^3\textbf {v}=-3\sum_{\beta,\zeta}\left(\frac{\mu_{\alpha \beta}}{m_\alpha+m_\beta}\right)\frac{n_{\alpha Z}}{\tau^{(Zh)}_{\alpha Z \beta \zeta}}(T_{\alpha Z}-T_{\beta \zeta})
\end{align}
where summation is performed over all species and time between collisions definition can be found in appendix \ref{app:1}. One can recognize a heat source due to friction between different ions in the second term of \eqref{eq:Q_diff}.

The friction term $\textbf R_{\alpha Z}$ defined according to \eqref{eq:def_R} does not need correction provided that Coulomb collisions do not lead to particle sources and sinks and three possible definitions are equivalent:
\begin{align}
m_\alpha \iiint (\textbf v - \textbf u_{\alpha Z}) C_{\alpha Z }d^3\textbf {v} = m_\alpha \iiint (\textbf v - \textbf u)C_{\alpha Z }d^3\textbf {v} = \textbf R_{\alpha Z}
\end{align} 

Therefore the friction term $\textbf R_{\alpha Z}$  found using the approach discussed in  \cite{book} may be directly substituted into the system \eqref{eq:density_eq} - \eqref{eq:ion_heat_eq}.

Note, according to Eq. (8.1.3) in \cite{book}, that momentum conservation in collisions is maintained:
\begin{align}
\sum_{\alpha, Z}\textbf R_{\alpha Z}=0
\end{align} 
So is energy conservation according to \eqref{eq:Q_diff}:
\begin{align}
\sum_{\alpha, Z}Q^{(Br)}_{\alpha Z}=-\sum_{\alpha, Z}\textbf u_{\alpha Z}\textbf R_{\alpha Z}
\end{align} 
Details about the conservative property of collisions can be found, for instance, in \cite{braginskii1965transport}.

Thus, assuming that the 21N-moment method is applied and transport coefficients are found, we need then to switch from $\textbf h_{\alpha Z}$ and $\overleftrightarrow{\pi}_{\alpha Z}$ to
$\textbf q^{(Br)}_{\alpha Z}$ and $\overleftrightarrow{\pi}^{(Br)}_{\alpha Z}$ by using \eqref{eq:hf_diff}, \eqref{eq:pi_diff} and taking into account the connection between temperatures \eqref{eq:temp_diff}.

In the present paper we consider only so-called parallel (with regard to the B-field) transport coefficients. The classical transport across the magnetic field is usually of little consequence for magnetic fusion devices, since the anomalous transport is in most cases much larger.

Then, as it is shown in \cite{book}, the transport coefficients appearing in the heat flux and friction force may be calculated independently of the viscosity coefficients. Therefore we may consider their corresponding calculations separately.

Finally, this approach can be applied both for electrons and ions. However, due to the small electron-ion mass ratio, transport for electrons can be considered separately, which significantly simplifies the approach. It is discussed in detail in \cite{book}. Consequently, all the analysis in this paper will be devoted to ion transport coefficients.

\subsection{Heat flux and friction term}

Applying the Grad method with 21N-moment (see \cite{book}), one can express $\textbf h_{\alpha Z}$ through the velocities $\textbf w_{\alpha Z}$ and temperature gradients $\nabla T_{\alpha Z}$ with the help of kinetic coefficients that can be found solving the algebraic system equations (8.4.2) in \cite{book}. For the parallel (with regard to the B-field) component
of $\textbf h_{\alpha Z}$, one gets (see Eq. (8.4.6) of \cite{book})):

\begin{multline}\label{eq:heat_flux_expression_par}
h_{\alpha Z \parallel}=
-\frac{n_{\alpha Z}}{n_{\alpha}}\sum_{\beta}\Big[\tilde {c}^{(h_T^A)}_{\beta\alpha}\frac{n_\beta T}{m_\beta}\tau^{(Zh)}_{\beta \alpha} \widetilde{\nabla_\parallel T_{\beta}}\Big]
-2{c}^{(h_T^B)}_{\alpha}\frac{n_{\alpha Z} T}{m_\alpha}\tau^{(Zh)}_{\alpha \alpha}\frac{\overline{Z_\alpha^2}}{Z^2}\nabla_\parallel T_{\alpha Z}\\
+n_{\alpha Z} T\sum_{\beta}{c}^{(h_w)}_{\beta \alpha}(w_{\alpha Z\parallel}-\overline{w}_{\beta \parallel})
\end{multline}
where $\displaystyle \nabla_\parallel = \left(\textbf b \cdot \nabla\right)$, and summation is performed over all types of ions and:
\begin{align}
\widetilde{\nabla_\parallel T_{\alpha}}=\sum_Z\frac{n_{\alpha Z}}{n_{\alpha}}\nabla_\parallel T_{\alpha Z}.
\end{align}
Collisional times $\tau^{(Zh)}_{\beta \alpha}$ and average squared charge $\overline{Z_\alpha^2}$ can be found in appendix \ref{app:1}. Kinetic coefficients $\tilde {c}^{(h_T^A)}_{ \beta\alpha}$, ${c}^{(h_w)}_{\beta \alpha}$ and  ${c}^{(h_T^B)}_{\alpha}$ are the result of solving the system  of algebraic equations (8.4.2) in \cite{book} and application of corrections for each charge state (details can be found in appendix  \ref{app:1}).

Note that the first two terms on the l.h.s. of \eqref{eq:heat_flux_expression_par} represent the thermal conductivity.  The third term is the velocity difference dependent part of the heat flux that, in the Braginskii approach \cite{braginskii1965transport}, only appears in the electron heat flux.

Then, following the method in \cite{book}, the friction term can be written in the same form as Eq.\ (8.4.5) of \cite{book}:
\begin{multline}\label{eq:Friction_term_Zh}
R_{\alpha Z\parallel}=
-n_{\alpha Z}\frac{Z^2}{\overline{Z_\alpha^2}}\sum_{\beta}\tilde{c}^{(R_T^A)}_{\alpha\beta}\widetilde{\nabla_\parallel T_{\beta}}-n_{\alpha Z}c^{(R_T^B)}_{\alpha}\nabla_\parallel T_{\alpha Z}\\
-n_{\alpha Z}\frac{Z^2}{\overline{Z_\alpha^2}}\sum_{\beta}\frac{\mu_{\alpha \beta}}{\tau^{(Zh)}_{\alpha \beta}}c^{(R_w)}_{\beta\alpha}(w_{\alpha Z\parallel}-\overline{w}_{\beta \parallel})
\end{multline}
where coefficients $\tilde {c}^{(R_T^A)}_{ \beta\alpha}$, ${c}^{(R_w)}_{\beta \alpha}$ and ${c}^{(R_T^B)}_{\alpha}$ are the result of solving the system  of algebraic equations (8.4.2) in \cite{book} and application of corrections for each charge state as well. In \eqref{eq:Friction_term_Zh}, the first two terms represent the thermal force, and the third term is the interspecies friction force.

\subsection{Viscosity}

Now let us consider the system of equations for viscosity. Note that, to describe parallel transport correctly  \cite[and references therein]{helander2005collisional}, it is required to take into account the heat flux dependence for the viscosity. However, this effect was not considered in the viscosity equations (8.1.6), (8.1.6`) in \cite{book}. The necessity to account for this heat viscosity requires a modification of the approach described in \cite{book}, namely,  using the general expression for the moment equation (A1.8) in \cite{book} and adding the heat flux dependent terms to the left-hand side of (8.1.6), (8.1.6`) in \cite{book} (after summation over charge states):

\begin{align}\label{eq:aver_viscosity_long_add}
p_{a}W_{\parallel\parallel}+W^{\overline h_{a}}_{\parallel\parallel}=\overline{R}^{20}_{\alpha\parallel\parallel}
\end{align}
where $\overline{R}^{20}_{\alpha\parallel\parallel}$ is the collisional right-hand side of Eq. (8.1.6) in \cite{book} summed over charge states, that depends on $\overline\pi_{\beta \parallel\parallel}$ ($\beta$ represents each ion in the mixture), and
\begin{align}\label{eq:aver_sigma_moment_long_add}
\frac{7}{2}\frac{T}{m_\alpha}W^{\overline h_{a}}_{\parallel\parallel}=\overline{R}^{21}_{\alpha\parallel\parallel}
\end{align}
where $\overline{R}^{21}_{\alpha\parallel\parallel}$ is the collisional right-hand side of Eq. (8.1.6`) in \cite{book} summed over charge states, that also depends on $\overline\pi_{\beta \parallel\parallel}$.

W-tensors and collisional right-hand sides can be found in appendix \ref{app:1}. The solution of this system of equations \eqref{eq:aver_viscosity_long_add}-\eqref{eq:aver_sigma_moment_long_add} can be expressed as:
\begin{multline}\label{eq:viscosity_result}
\pi_{\alpha Z \parallel\parallel}=-\frac{n_{\alpha Z}}{n_{\alpha}}\left(\sum_{\beta}\big[\tilde{c}^{(\pi_u^A)}_{\alpha\beta}+2c^{(\pi_u^B)}_{\alpha}\frac{\overline{Z_\alpha^2}}{Z^2}\delta_{\alpha\beta}\big]\tau^{(Zh)}_{\beta \alpha}p_{\beta}\right)W_{\parallel\parallel}\\
-\frac{n_{\alpha Z}}{n_{\alpha}}\sum_{\beta}\left(\tilde{c}^{(\pi^A_h)}_{\alpha\beta}\tau^{(Zh)}_{\beta \alpha}W^{\overline h_{\beta}}_{\parallel\parallel}
\right)-2c^{(\pi^B_h)}_{\alpha}\frac{\overline{Z_\alpha^2}}{Z^2}\tau^{(Zh)}_{\alpha \alpha}W^{h_{\alpha Z}}_{\parallel\parallel}
\end{multline}
where dimensionless transport coefficients $\tilde c^{(\pi_u^A)}_{\alpha \beta}, \tilde c^{(\pi_h^A)}_{\alpha \beta}$ are the result of solving the system of algebraic equations \eqref{eq:aver_viscosity_long_add} and \eqref{eq:aver_sigma_moment_long_add}, and the corrections for each charge state $c^{(\pi_u^B)}_{\alpha}, c^{(\pi_h^B)}_{\alpha}$ can be found analytically, as it is done for the heat flux (appendix \ref{app:1}). The first term in \eqref{eq:viscosity_result} is the velocity dependent part of the viscosity that is discussed in \cite{book}. The last terms in  \eqref{eq:viscosity_result} represent additional effects due to the heat flux that is taken into account in this paper by adding heat flux dependent terms into the left-hand side both in \eqref{eq:aver_viscosity_long_add} and \eqref{eq:aver_sigma_moment_long_add}.

Usually, it is not a trivial task in complex geometry to take a divergence $\nabla \cdot \overleftrightarrow{\pi}_{\alpha Z}$, intended to be included into \eqref{eq:momentum_eq}, even in case \eqref{eq:viscosity_result}. One way to do it supposes to take into account only parallel and drift components of the velocity and parallel and diamagnetic components of the heat flux in the viscosity calculation. Details can be found in appendix \ref{app:2}.

Finally, to apply the closure discussed in \cite{book} and in this paper for the Braginskii system of equations \eqref{eq:density_eq}-\eqref{eq:ion_heat_eq}, corrections of temperatures,
heat fluxes and viscosities should be done according to \eqref{eq:temp_diff}, \eqref{eq:hf_diff} and \eqref{eq:pi_diff}.

Now, let us recall results of transport in the Pfirsch-Schlüter regime \cite{helander2005collisional,Hirshman_1981}. For a simple plasma, using the moment approach, parallel viscosity transport coefficients (in the Pfirsch-Schlüter regime) were found in \cite{Hirshman_1981}.
In the single ion species case, one can calculate viscosity transport coefficients using \eqref{eq:viscosity_result} as well. Thus, in the variables used in \cite{helander2005collisional} they are:
\begin{align}\label{eq:mu_alpha_1_2}
\mu_{\alpha1}=p_{i}2\tau^{(Zh)}_{ii}\times0.96,\ \ \ \mu_{\alpha2}=p_{i}2\tau^{(Zh)}_{ii}\times1.55
\end{align}
That is close to what was obtained  in \cite{helander2005collisional,Hirshman_1981} and applied in \cite{Rozhansky_2001, Rozhansky_2009}. Therefore, in the collisional case, it is expected to obtain a radial electrical field close to the radial electrical field in the Pfirsch-Schlüter regime \cite{Hinton1976}.
It is important to mention that to get $\mu_{\alpha2}$ as in \eqref{eq:mu_alpha_1_2}, it is necessary to add the heat flux dependent term both into \eqref{eq:aver_viscosity_long_add} and \eqref{eq:aver_sigma_moment_long_add}. As a result, using this approach, the result \eqref{eq:mu_alpha_1_2} is extended to arbitrary plasma mixtures.

\subsection{Summary}

In this section we have considered an explicit method based on solving an algebraic system of equations, which can be applied for the various plasma compositions, for example deuterium, tritium, helium, and other impurities. This case occurs in current devices \cite{Litaudon_2017} and will be standard operating procedure in future reactors \cite{Ikeda_2007,FEDERICI2014882,Wan2017}. Thus it should be implemented into codes like SOLPS-ITER \cite{WIESEN2015480,Bonnin2016} that are used for fusion reactor operation predictions. Section \ref{For_SOlPS} is dedicated to this.

In the next section, an analytical approach is considered  that can be applied for many cases which is, however, less accurate for species with close masses.

\section{Improved analytical expressions}\label{section:Analyt}

\subsection{Heat flux}\label{subsection:heat_flux}

Let us consider the closure in the case with one light and several heavy ion species. The resulting $2n_s\times2n_s$ matrix (where $n_s$ is the number of different types of species (which have different atomic nucleus) in a mixture) which is intended to be inverted to solve system of equations (8.4.2) \cite{book}, can be split into blocks:
\begin{align}
A=
\begin{bmatrix}
A_{11} & A_{12}\\
A_{21} & A_{22}
\end{bmatrix}; \ \ \ \
A^{-1}=
\begin{bmatrix}
\tilde A_{11} & \tilde A_{12}\\
\tilde A_{21} & \tilde A_{22}
\end{bmatrix};
\end{align}
where each block can be written:
\begin{align}
A_{qp}=
\begin{bmatrix}
a^{qp}_{00} & a^{qp}_{0 1} & \cdots & a^{qp}_{0 n_s}\\
a^{qp}_{10} & a^{qp}_{11} & \cdots & a^{qp}_{1n_s}\\
\vdots & \vdots & \ddots & \vdots \\
a^{qp}_{n_s0} & a^{qp}_{n_s1} & \cdots& a^{qp}_{n_sn_s}
\end{bmatrix}; \ \ \ \
\tilde A_{qp}=
\begin{bmatrix}\label{eq:martix}
\tilde a^{qp}_{00} & \tilde a^{qp}_{0 1} & \cdots & \tilde a^{qp}_{0 n_s}\\
\tilde a^{qp}_{10} & \tilde a^{qp}_{11} & \cdots & \tilde a^{qp}_{1n_s}\\
\vdots & \vdots & \ddots & \vdots \\
\tilde a^{qp}_{n_s0} & \tilde a^{qp}_{n_s1} & \cdots& \tilde a^{qp}_{n_sn_s}
\end{bmatrix}
 \ \ \ \
\end{align}
Here and further index $0$ represents the light species, while indices $1..n_s$ correspond to heavy species. Indices $q=1,2$ and $p=1,2$ define corresponding blocks. 
Note that in the codes, which are applied for light main ion and heavy impurities case modeling \cite{WIESEN2015480}, main and impurity species are usually explicitly specified, and equations for them are written differently. The approach discussed in this paper allow us to write equations uniformly for all species.

First of all, for the  diagonal elements, estimations can be made ($i\neq0$):
\begin{align}\label{eq:estimations_1}
\frac{a^{qp}_{00}}{ a^{qp}_{0i}}\propto\frac{m_{i}}{ m_{0}}\sum_{\beta=0..n_s}\frac{\overline{Z_\beta^2}n_\beta}{\overline{Z_i^2}n_i}
\end{align}
\begin{align}
A_{11}, A_{12}: {a^{1p}_{0i}}={ a^{1p}_{i0}};
\ \ \ A_{21}, A_{22}: \frac{a^{2p}_{0i}}{ a^{2p}_{i0}}\propto\frac{m_{i}}{ m_{0}}
\end{align}
Taking into account, that  $\sum_{\beta=0..n_s}\frac{\overline{Z_\beta^2}n_\beta}{\overline{Z_i^2}n_i}\geq1$, cross elements in the first row and column are smaller than $a^{qp}_{00}$ by at least the mass ratio. Thus, $\tilde a^{qp}_{00}$ elements of the inverted matrix can be obtained independently from the contributions due to heavy species in the cross elements and $\tilde a^{qp}_{00} \gg \tilde a^{qp}_{0i}$.

Consider the solution for the temperature dependent part of the heat flux (heat conductivity):
\begin{align}\label{eq:aver_heat_flux_expression_1}
\frac{\overline{\textbf h}^T_{\alpha}}{p_\alpha}=
\frac{5}{2}\sum_{\beta}\tilde { {a}}^{11}_{\alpha\beta}n_{\beta}\widetilde{\nabla T_{\beta}}
\end{align}
where $\tilde { {a}}^{11}_{\alpha\beta}$ is an element from the $\tilde A_{11}$ part of the inverted matrix. In the most common case, where the density of the light species is higher than the density of the heavy species, the impact from cross elements on the light species heat flux is even smaller due to multiplication with $n_{\beta}$. On the other hand, for the heavy species heat flux all terms in the sum are comparable; however it is easy to show that the heavy species heat flux in such case does not contribute significantly to the global heat balance and therefore $T_{0}$ temperature for light species averaged over all charge states. Temperature of the light species $T_{0}$ plays major role in the thermal force between light and heavy species (see subsection \ref{Friction_sec}), and thus, determines heavy impurity transport.

Thus, in the case with one light and several heavy species, the transport coefficients can be obtained by solving equations for each type of species independently, ignoring cross elements. This allows us to derive analytical expressions for the transport coefficients (see appendix \ref{app:1}).

These expressions are an improved form of the Zhdanov analytical expressions (8.4.7) in \cite{book}. Zhdanov suggests obtaining transport coefficients keeping only terms with $(m_0/m_i)^0$ order, which directly exclude cross elements from consideration (due to \eqref{eq:estimations_1}), while the approach discussed in this paper keeps higher order terms $(m_0/m_i)^n$, where $n=1..\infty$ in diagonal elements. The reason for this action follows from the fact that the diagonal elements play a major role. Analysis of the matrix in the test cases confirms this fact: even for the mixtures, where masses of species become close, cross elements were smaller than diagonal elements.
Therefore, higher order terms in diagonal elements affect the answer more than for the cross terms.

\begin{figure}
  \centering
\includegraphics[scale=0.3]{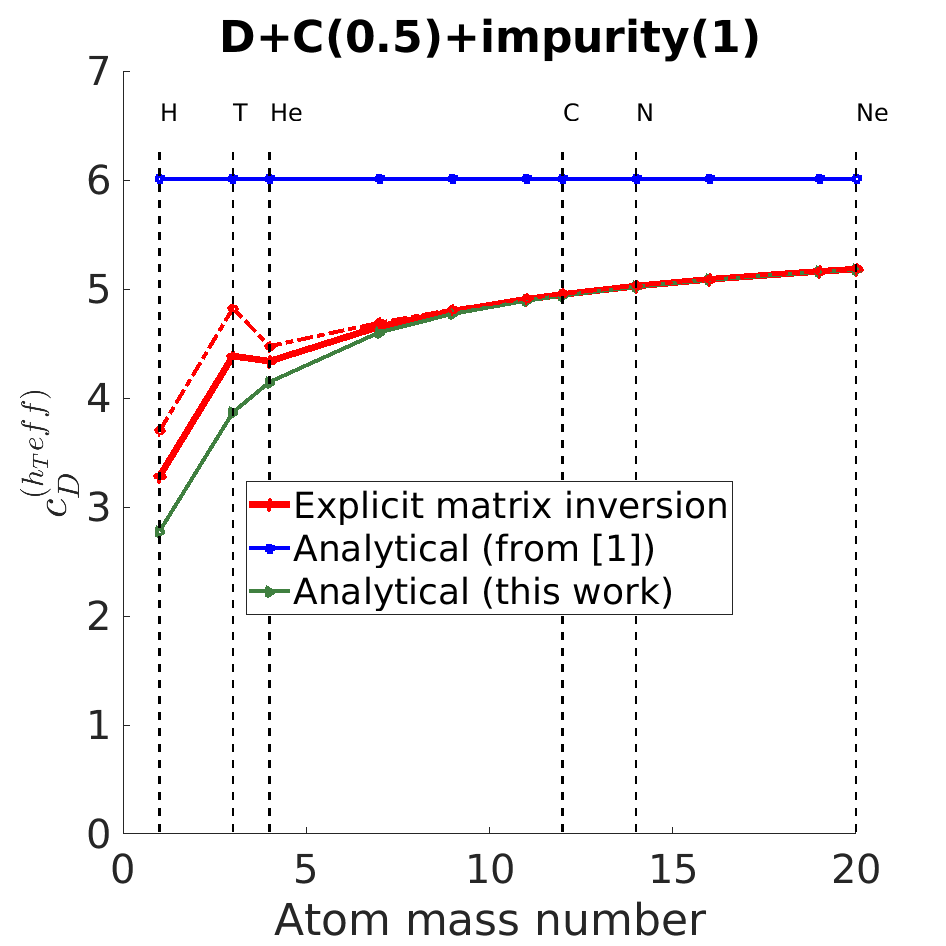}  
\caption{Temperature dependent part of the heat flux transport coefficient for deuterium using the explicit solution of the system of equations (8.4.2) in \cite{book}, the original Zhdanov formula (8.4.7) in \cite{book} and the improved formula (appendix \ref{app:1}) for the D + C + another impurity cases with equal distribution between charge states. \\
The additional impurity is varied along the horizontal axis.\\
The dash-dotted line shows the case where: $\widetilde{\nabla_{\parallel} T_{i}}=2\times \nabla_{\parallel} T_{D}$.\\
The number in parentheses at the top of the figure gives normalised impurity density ($={\overline{Z_i^2}n_i}/{n_D}$ where $\overline{Z_i^2}$ is an averaged square of the charge of impurity defined in \eqref{eq:Z2_def}) and is kept constant. }
\label{fig:c_63_D_C_imp}  
\end{figure}

To prove this assumption, let us compare the heat conductivity for  deuterium in the presence of impurity ions found using this approach with the result using explicit matrix inversion (for solving algebraic system equations (8.4.2) in \cite{book}) and the Zhdanov analytical expression (8.4.7) in \cite{book}. Consider the case (deuterium, carbon and another impurity), where the temperature gradient for both impurities is the same as the temperature gradient for deuterium, and the case, where the temperature gradient for both impurities is (for some reason) 2 times higher than for deuterium, to explore what role the difference in ion temperatures plays. In test cases considered in this paper, the amount of carbon and another impurity is chosen according to the rules ${\overline{Z_C^2}n_C}/{n_D}=0.5$, $ {\overline{Z_I^2}n_I}/{n_D}=1.0$, where "I" corresponds to another impurity and $\overline{Z_i^2}$ is the averaged square of the impurity charge defined in \eqref{eq:Z2_def}.
The heat flux associated with the deuterium heat conductivity is:
\begin{align}
h^{T}_{D\parallel}=-{c}^{(h_T{eff})}_{D}\frac{n_DT\tau^{(Zh)}_{D}}{m_D}\nabla_{\parallel} T_{D}
\end{align}
where the transport coefficient ${c}^{(h_T{eff})}_{D}$ is shown in Figure~\ref{fig:c_63_D_C_imp} and the collision time is defined in appendix \ref{app:1}.

In Figure~\ref{fig:c_63_D_C_imp} different results are plotted: with similar masses and densities for light impurities and small ratios $m_D/m_{imp}$ and $n_{imp}/n_D$ for heavy impurities.

\begin{figure}
  \centering
\includegraphics[scale=0.3]{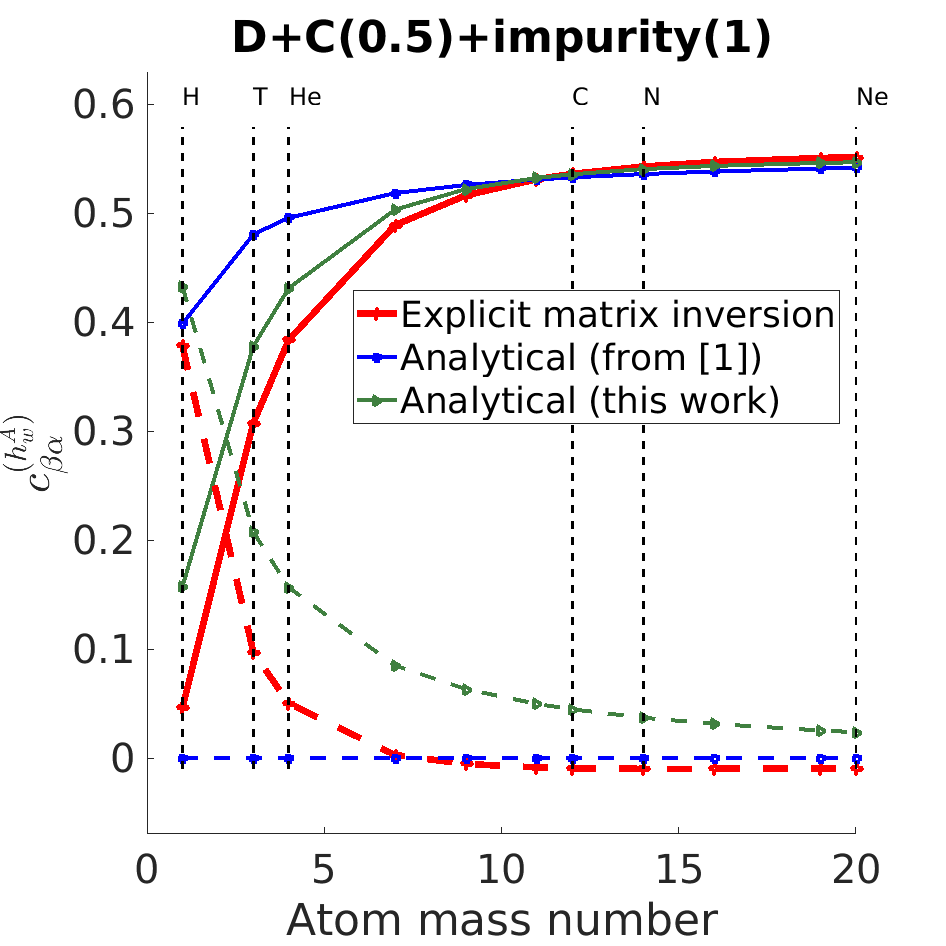}  
\caption{Velocity dependent part of the heat flux transport coefficient for the deuterium due to D/impurity velocity difference (solid) and for another impurity due to impurity/D velocity difference (dashed) using explicit solution of the system of equations (8.4.2) in \cite{book}, the original Zhdanov formula (8.4.7) in \cite{book} and our improved formula (appendix \ref{app:1}) for  D+C+another impurity. (See the caption to Figure \ref{fig:c_63_D_C_imp} for additional information.)
}
\label{fig:c_2_D_C_imp}  
\end{figure}

It is clearly shown that $(m_0/m_i)^0$ order accuracy is not sufficient to get good agreement with the explicit matrix inversion result, while the addition of higher order terms only in the diagonal elements provides solutions close to the numerical method except in the hydrogen-helium region, where cross elements in the matrix become important. In this region, a different temperature gradient for the impurity affects the result because both close masses and densities in the impurity term in  \eqref{eq:aver_heat_flux_expression_1} play a role.

The same comparison can be made for the velocity dependent part of the heat flux:
\begin{align}
\overline{h}_{\alpha \parallel}^{w}=
p_{\alpha}\sum_{\beta}c^{(h_w^A)}_{\beta \alpha}(\overline{w}_{\alpha \parallel}-\overline{w}_{\beta \parallel})
\end{align}
where the transport coefficient $c^{(h_w^A)}_{\beta \alpha}$ is shown in Figure~\ref{fig:c_2_D_C_imp}.

The velocity dependent part of the deuterium heat flux in the presence of high mass impurities predicted by both original (8.4.7) in \cite{book} and improved analytical expressions is close to the result obtained from the matrix inversion approach. However, for impurities lighter than beryllium, the improved formula gives a solution much closer to the numerical one (Figure~\ref{fig:c_2_D_C_imp}) than the Zhdanov formula. On the other hand, for the high mass species, the  calculated heat flux is less accurate (Figure~\ref{fig:c_2_D_C_imp}). In the original monograph \cite{book}, it is suggested to set to zero transport coefficients that represent heavy-light species interactions in the heat flux (velocity dependent part) for heavy species. Although that the improved formula \eqref{eq:heat_flux_vel_aver_coef} in appendix \ref{app:1} suppresses the coefficient by a factor $(\mu_{\alpha \beta}/m_\alpha)^{3/2}$ for the cases, where $m_\alpha>m_\beta$, numerical calculation provides even smaller velocity dependent part of the heat flux (Figure~\ref{fig:c_2_D_C_imp}). In the extreme case, where a difference between masses of species is large, for instance ion-electron mixture in the simple plasma, the velocity dependent part is set equal to zero for heavy species \cite{book,braginskii1965transport}. Thus, the reduction of the velocity dependent part for heavy species can be implemented by setting \eqref{eq:heat_flux_vel_aver_coef} equal to zero if $m_\alpha/m_\beta<1$. To keep all expressions the same for both low and high mass species, smooth analytical approximation to the step function can be applied.

Finally, following \cite{book}, the heat flux for each charge state can be found (appendix \ref{app:1}).

\subsection{Friction term}\label{Friction_sec}

Now consider the friction term. According to (8.1.3) \cite{book}:
\begin{multline} \label{eq:aver_Friction_term_expression_1}
{\textbf{R}_{\alpha}}\equiv\sum_Z{\textbf{R}_{\alpha Z}}=\sum_{\beta}\Big[ \overline G^{(1)}_{\alpha \beta }(\overline{\textbf w}_{\alpha}-\overline{\textbf w}_{\beta })+\frac{\mu_{\alpha \beta}}{T}\overline G^{(2)}_{\alpha \beta }\left(\frac{\overline{\textbf h}_{\alpha}}{m_{\alpha}n_{\alpha}}-\frac{\overline{\textbf h}_{\beta }}{m_{\beta}n_{\beta}}\right)\\
+\left(\frac{\mu_{\alpha \beta}}{T}\right)^2\overline G^{(8)}_{\alpha \beta }\left(\frac{\overline{\textbf r}_{\alpha}}{m_{\alpha}n_{\alpha}}-\frac{\overline{\textbf r}_{\beta }}{m_{\beta}n_{\beta}}\right)\Big]
\end{multline}
According to \eqref{eq:aver_Friction_term_expression_1}, one can conclude that $\sum_\alpha{\textbf{R}_{\alpha}}=0$ for any velocities,  heat fluxes and r-moments (5-order vector moments see Eq. (8.1.2) in \cite{book}) ($\overline G^{(1)}_{\alpha \beta },\overline G^{(2)}_{\alpha \beta },\overline G^{(8)}_{\alpha \beta }$ are symmetric regarding to the replacement of $\alpha\beta$ with $\beta\alpha$).

Note that the impact on the friction term of species $\alpha$ from the heat flux and r-moment of species $\beta$ is proportional to ${\mu_{\alpha \beta}}/{m_\beta}$ and $\left({\mu_{\alpha \beta}}/{m_\beta}\right)^2$ correspondingly. Hence, leading terms in both ${\textbf{R}_{0}}$ and ${\textbf{R}_{i}}$, where $i\neq0$, are dependent on the heat flux and r-moment of the light species and, therefore, on $T_{0}$. Thus, where the heat flux/r-moment of heavy species are calculated inaccurately, it does not affect the friction term for heavy species (and light as well) summed over all charge states.

The friction term for each charge state can be found using:
\begin{multline} \label{eq:Friction_diff}
\textbf{R}_{\alpha Z}=I_{\alpha Z}{\textbf{R}_{\alpha}}+I_{\alpha Z}\sum_{\beta}\Big[ \overline G^{(1)}_{\alpha \beta }(\textbf w_{\alpha Z}-\overline{\textbf w}_{\alpha})\\
+\frac{\mu_{\alpha \beta}}{m_{\alpha}}\overline G^{(2)}_{\alpha \beta }\left(\frac{\textbf h_{\alpha Z}}{p_{\alpha Z}}-\frac{\overline{\textbf h}_{\alpha}}{p_{\alpha}}\right)
+\left(\frac{\mu_{\alpha \beta}}{m_{\alpha}}\right)^2\frac{m_{\alpha}}{T}\overline G^{(8)}_{\alpha \beta }\left(\frac{\textbf r_{\alpha Z}}{p_{\alpha Z}}-\frac{\overline{\textbf r}_{\alpha}}{p_{\alpha}}\right)\Big]
\end{multline}
Substituting results of the heat flux and r-moment into \eqref{eq:aver_Friction_term_expression_1} and \eqref{eq:Friction_diff}, one can obtain analytical transport coefficients for the thermal and friction forces (appendix \ref{app:1}).

Now we compare these results with the explicit matrix inversion method and the coefficients previously implemented into SOLPS-ITER \cite{Sytova2018,Sytova2020}, which are based on the Zhdanov analytical expressions (8.4.7) in \cite{book}.

\begin{figure}
  \centering
\includegraphics[scale=0.3]{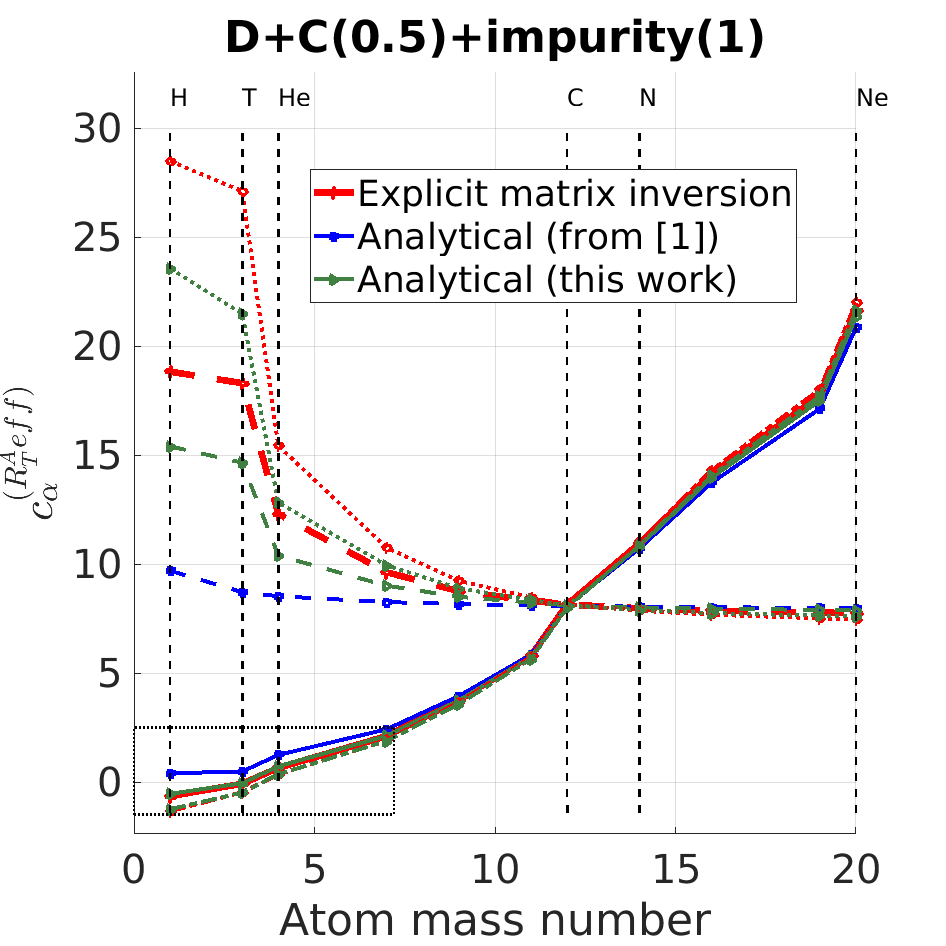}  
\caption{Temperature dependent part of the friction term transport coefficient for another impurity (solid) and  for the carbon (dashed) using explicit solution of the system of equations (8.4.2) in \cite{book}, the original formula from \cite{Sytova2018,Sytova2020} and our improved formula (appendix \ref{app:1}) for D+C+another impurity. 
Dash-dotted (another impurity) and dotted (carbon) lines are plotted for the case where $\widetilde{\nabla_{\parallel} T_{i}}=2\times \nabla_{\parallel} T_{D}$.  
(See the caption to Figure \ref{fig:c_63_D_C_imp} for additional information.)
}
\label{fig:Omega_D_C_i_Zeff_corr}  
\end{figure}

\begin{figure}
  \centering
\includegraphics[scale=0.3]{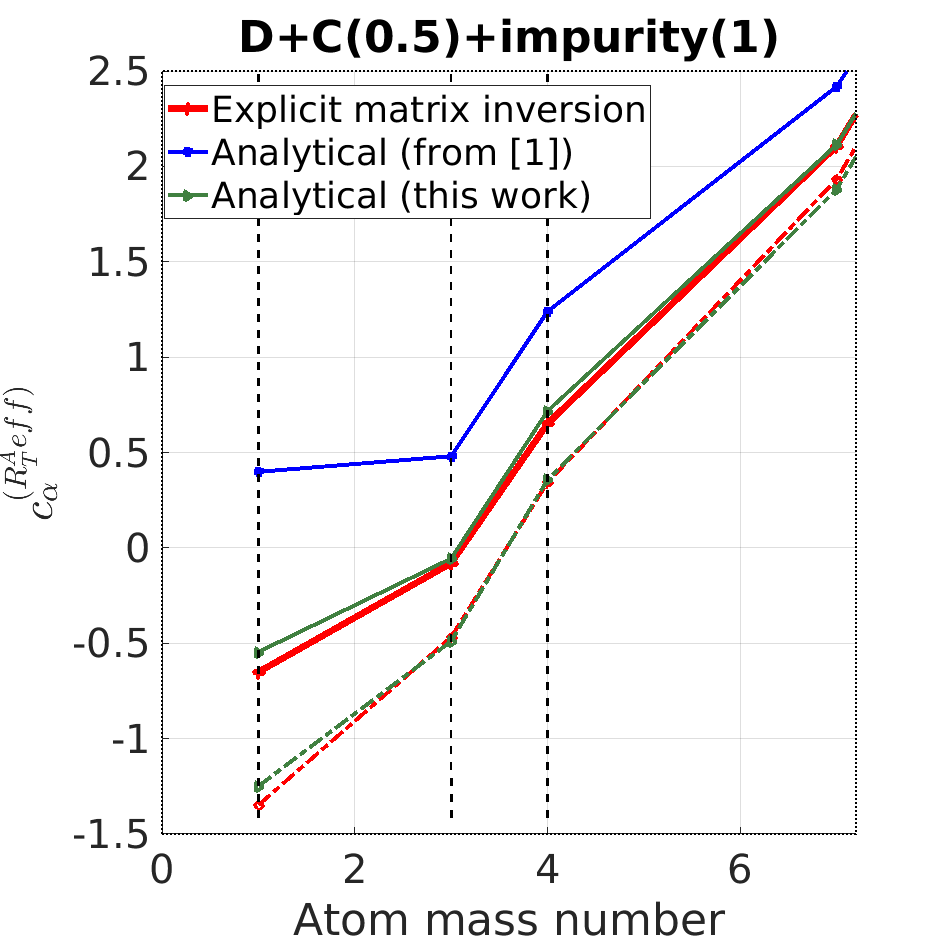}  
\caption{Zoom in the dotted box of the Figure~\ref{fig:Omega_D_C_i_Zeff_corr}
}
\label{fig:Omega_D_C_i_Zeff_corr_Zoom_in}  
\end{figure}

First of all, note that, in the trace-impurity case (${\overline{Z_i^2}n_i}/{\overline{Z_0^2}n_0}\ll1$, $m_0<m_i$ ), transport coefficients in \cite{Sytova2018,Sytova2020} for the friction force yield $c^{(1)}_{imp}\approx1$ and, for the thermal force, $c^{(2)}_{imp}\approx1.56$. However, according to  \eqref{eq:aver_Friction_term_expression_1}, the thermal force depends on the masses of the participants (besides the factor $\sqrt{\mu_{\alpha\beta}}$ in the G-matrices). Indeed, formulas \eqref{eq:fric_force_aver} and \eqref{eq:fr_term_5} in appendix \ref{app:1}, using the notation from \cite{Sytova2018,Sytova2020}, provide:
\begin{align}\label{eq:tran_coef_trace}
c^{(1)}_{imp}\approx1,\ \ \ c^{(2)}_{imp}\approx1.56\frac{\mu_{0 i}}{m_{0}}\Big[\frac{3}{2}-\frac{1}{2}\frac{\mu_{0 i}}{m_0}\Big]
\end{align}
(here the second term in \eqref{eq:fr_term_5} for species 0 is absent, which becomes important if $m_0\approx m_i$). Thus, the light trace-impurity thermal force in \cite{Sytova2018,Sytova2020} is larger than our improved expressions.

Then, consider the thermal force for each type of species summed over all charge states in the non-trace-impurity case for a case (deuterium, carbon and another impurity) where the temperature gradient for both impurities is the same as the temperature gradient for deuterium, and  the case, where the temperature gradient for both impurities is (for some reason) 2 times larger than for deuterium, to find where difference in ions temperatures play a role. The transport coefficient for the thermal force averaged over all charge states for carbon and another impurity (Figures~\ref{fig:Omega_D_C_i_Zeff_corr},~\ref{fig:Omega_D_C_i_Zeff_corr_Zoom_in}) is described by:
\begin{align}
R^{T}_{\alpha}=
n_{\alpha}c^{(R_T^{A}eff)}_{\alpha}\nabla_{\parallel} T_{D}
\end{align}

For a deuterium plasma with carbon, the original Zhdanov analytical formula results in a $\sim1\%$ deviation from the numerically calculated result. Therefore for carbon and heavier impurities this formula can be applied, although, a slight increase of deviation for the heavier impurity due to the presence of carbon is seen (Figure~\ref{fig:Omega_D_C_i_Zeff_corr}). Thus, the thermal force averaged over all charge states for nitrogen and neon calculated using the 3.0.6 SOLPS-ITER version and higher (and as a result in \cite{Kaveeva_2020,senichenkovstudy,SYTOVA201972,Sorokina2018} and others) is the same as can be obtained by numerically solving the system of equations for heat flux and r-moment (5-order vector moment) (8.4.2) in \cite{book}.

Furthermore, in the region of heavy impurities, the impurity temperature gradient does not play a role (Figure~\ref{fig:Omega_D_C_i_Zeff_corr}). Indeed, the shape of the impurity distribution function (represented by impurity heat flux and r-moment (5-order vector moment, see Eq. (8.1.2) in \cite{book}) in the Hermite polynomial expansion) is not important in the case where $m_D\ll m_i$, $T_D\approx T_i$, and thus $v_D\gg v_i$. Shaping of the deuterium distribution function driven by $ \nabla_{\parallel} T_{D}$ is a major effect in such cases.

However, in the light impurity region the situation is different. For cases where $\nabla_{\parallel} T_{D}=\widetilde{\nabla_{\parallel} T_{i}}$, the analytical Zhdanov expression results in an up to 15\% deviation in a (D+C+Li) mixture and up to 90\% deviation in a (D+C+He) mixture (Figure~\ref{fig:Omega_D_C_i_Zeff_corr_Zoom_in}). If the temperature of the impurity is different, the deviation is even larger. In contrast, the improved analytical expression for the thermal force in (D+C+Li) gives only a 1\% deviation for lithium (Figure~\ref{fig:Omega_D_C_i_Zeff_corr_Zoom_in}) and  6\% deviation for carbon (Figure~\ref{fig:Omega_D_C_i_Zeff_corr}) from the numerically calculated result. And for the (D+C+He) case, the improved formula provides up to 10\% deviation for helium (Figure~\ref{fig:Omega_D_C_i_Zeff_corr_Zoom_in}) and up to 16\% for carbon (Figure~\ref{fig:Omega_D_C_i_Zeff_corr}). Note that in this region, the temperature gradient of impurities becomes important.

Notice that deviations for carbon are due to the presence of significant amount of helium ($n_{He}/n_D=0.4$) and lithium ($n_{Li}/n_D=0.2$), for smaller amounts of helium/lithium this case is closer to the pure deuterium/carbon. In the ($n_{He}/n_D=0.04$) case, the deviation from the explicit matrix inversion result is up to 5\% for carbon, however becoming worse for helium: up to 30\% (up to 50\% for the case, where $\widetilde{\nabla_{\parallel} T_{He}}=2\times \nabla_{\parallel} T_{D}$).

For the cases: (deuterium, carbon and hydrogen); and (deuterium, carbon and tritium), the original formula (8.4.7) in \cite{book} provides even the wrong sign (always positive).  However, the new formula \eqref{eq:fr_term_5} in appendix \ref{app:1} works surprisingly well, which confirms the observation made in the tests that even for comparable masses, cross elements play a minor role. In mixtures (D+C+H/T). the thermal force has a deviation for hydrogen/deuterium/tritium of up to 30\% (up to 50\% for the case, where $\nabla_{\parallel} T_{T}=2\times \nabla_{\parallel} T_{D}$) and for carbon 15-20\% deviation from numerically calculated result (figures~\ref{fig:Omega_D_C_i_Zeff_corr},~\ref{fig:Omega_D_C_i_Zeff_corr_Zoom_in}). For more accurate calculations in case of comparable masses, the explicit solution of the system of equations (8.4.2) in \cite{book} is required. Moreover, for such cases the temperature gradient of impurities plays a role, and heat fluxes of impurities are important and an accurate calculation of these fluxes is also necessary.

\begin{figure}
  \centering
\includegraphics[scale=0.3]{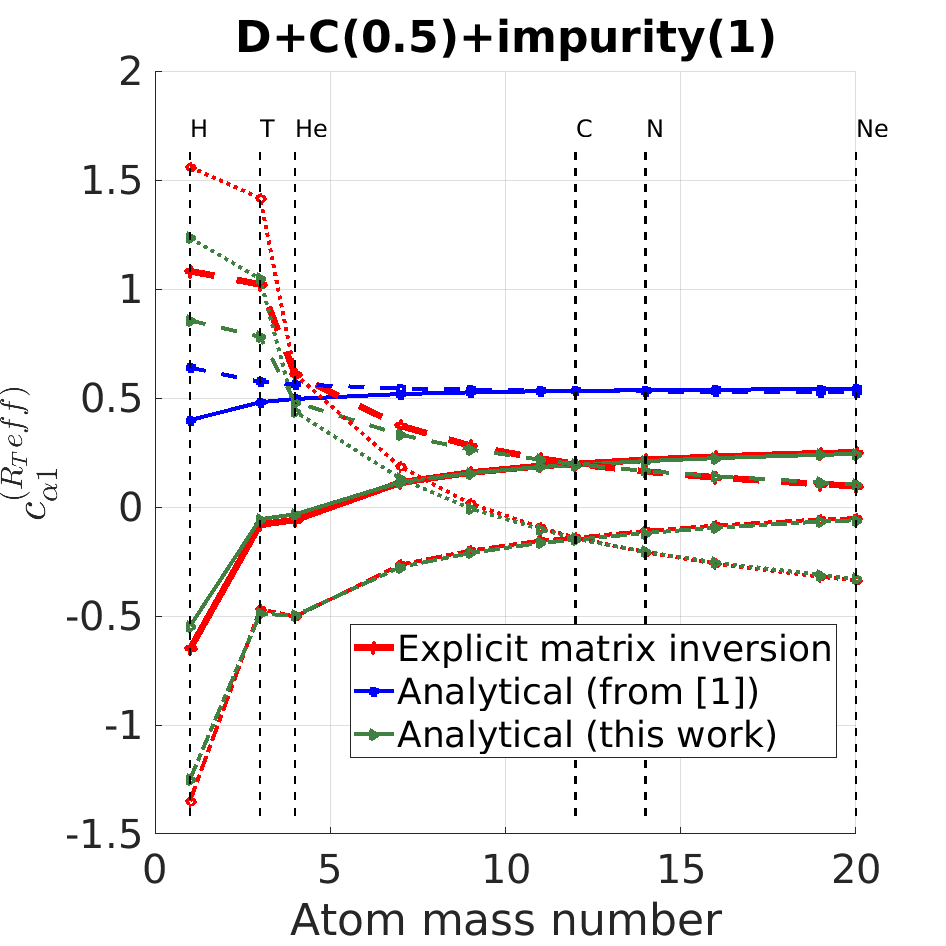}  
\caption{Temperature dependent part of the friction term transport coefficient for another impurity $Z=+1$ charge state (solid) and  for the carbon $Z=+1$ charge state (dashed) using the explicit solution of the system of equations (8.4.2) in \cite{book}, the original formula from \cite{Sytova2018,Sytova2020} and our improved formula (appendix \ref{app:1}) for D + C + another impurity.
Dash-dotted (another impurity $Z=+1$) and dotted (carbon $Z=+1$) lines show the case where $\widetilde{\nabla_{\parallel} T_{i}}=\nabla_{\parallel} T_{i1}=2\times \nabla_{\parallel} T_{D}$.
(See the caption to Figure \ref{fig:c_63_D_C_imp} for additional information.)
}
\label{fig:Omega_D_C_i_Z1}  
\end{figure}

Now consider the thermal force for each charge state species obtained using \eqref{eq:Friction_diff}, which affects space separation between different charge states of the same type of species. Note that this procedure does not require additional assumptions, thus can be made straightforwardly even for mixtures of species with close masses (appendix \ref{app:1}). What was done for the thermal force in \cite{Sytova2018,Sytova2020} represents what is done in the first term in \eqref{eq:Friction_diff}, therefore the dependence on the difference between heat flux and r-moment (5-order vector moment, see Eq. (8.1.2) in \cite{book}) averaged over charge states and for each charge state was not taken into account.

Here again consider the case where the temperature gradient for impurities is the same as the temperature gradient for deuterium, and the case, where temperature gradient for impurities (all charge states) is 2 times larger than for deuterium, to find, where the difference in ion temperatures plays a role. In Figure~\ref{fig:Omega_D_C_i_Z1}, the transport coefficient for the thermal force for the first charge state for carbon and another impurity described by:
\begin{align}
R^{T}_{\alpha1}=
n_{\alpha1}c^{(R_T{eff})}_{\alpha1}\nabla_{\parallel} T_{D}
\end{align}
is plotted.

According to the thermal force for each charge state, expression \eqref{eq:chst_Thermal_force} in appendix \ref{app:1}, that is written taking into account the correction to the average thermal force, the first charge state is affected by the correction more than the others, because the impact from the average thermal force is reduced by a factor $1/\overline{Z_\alpha^2}$.
It is clearly seen (Figure~\ref{fig:Omega_D_C_i_Z1}) that, for the first charge state of the impurity, the correction, represented by the third and fourth term in \eqref{eq:Friction_diff} (blue curves represent the model from \cite{Sytova2018,Sytova2020}), plays a role. As a result, the temperature gradient for impurities becomes important (Figure~\ref{fig:Omega_D_C_i_Z1}). However, for carbon and higher mass impurities, the thermal force that leads to space separation of each charge state, is an order of magnitude smaller than the thermal force that drives impurities, summed over all charge states (Figure~\ref{fig:Omega_D_C_i_Zeff_corr}). Therefore, it is expected that for the heavy impurities, their temperature gradient affects slightly the the charge states space distribution, but not the global impurity transport.

\begin{figure}
  \centering
\includegraphics[scale=0.3]{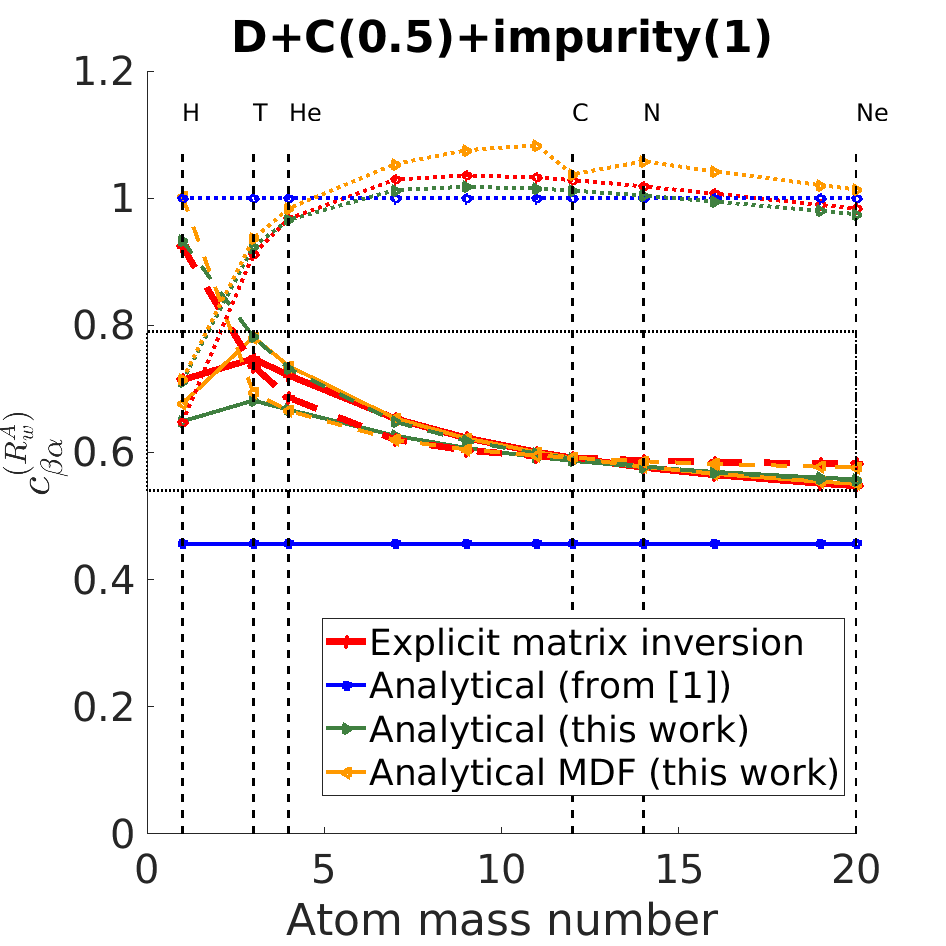}  
\caption{Velocity dependent part of the friction term transport coefficient for another impurity/deuterium (solid), for carbon/deuterium (dashed) and for  another impurity/carbon (dotted) using the explicit solution of the system of equations (8.4.2) in \cite{book}, the original formula from \cite{Sytova2018,Sytova2020}, and our improved formula (appendix \ref{app:1}) for the D + C + another impurity case. (See the caption to Figure \ref{fig:c_63_D_C_imp} for additional information.)
}
\label{fig:c_1_D_C_imp} 
\end{figure}

Then, consider the friction force for each type of species summed over all charge states. Take a look at the case (deuterium, carbon and another impurity). In Figure~\ref{fig:c_1_D_C_imp} are plotted transport coefficients for the friction force, averaged over all charge states, for carbon and another impurity, described by:

\begin{align}
R^{\textbf w}_{\alpha}=
-n_\alpha\sum_{\beta}\frac{\mu_{\alpha \beta}}{\tau^{(Zh)}_{\alpha \beta}}c^{(R_w^A)}_{\beta \alpha}(\overline{w}_{\alpha \parallel}-\overline{w}_{\beta \parallel})
\end{align}
where definitions can be found in appendix \ref{app:1}.

\begin{figure}
  \centering
\includegraphics[scale=0.3]{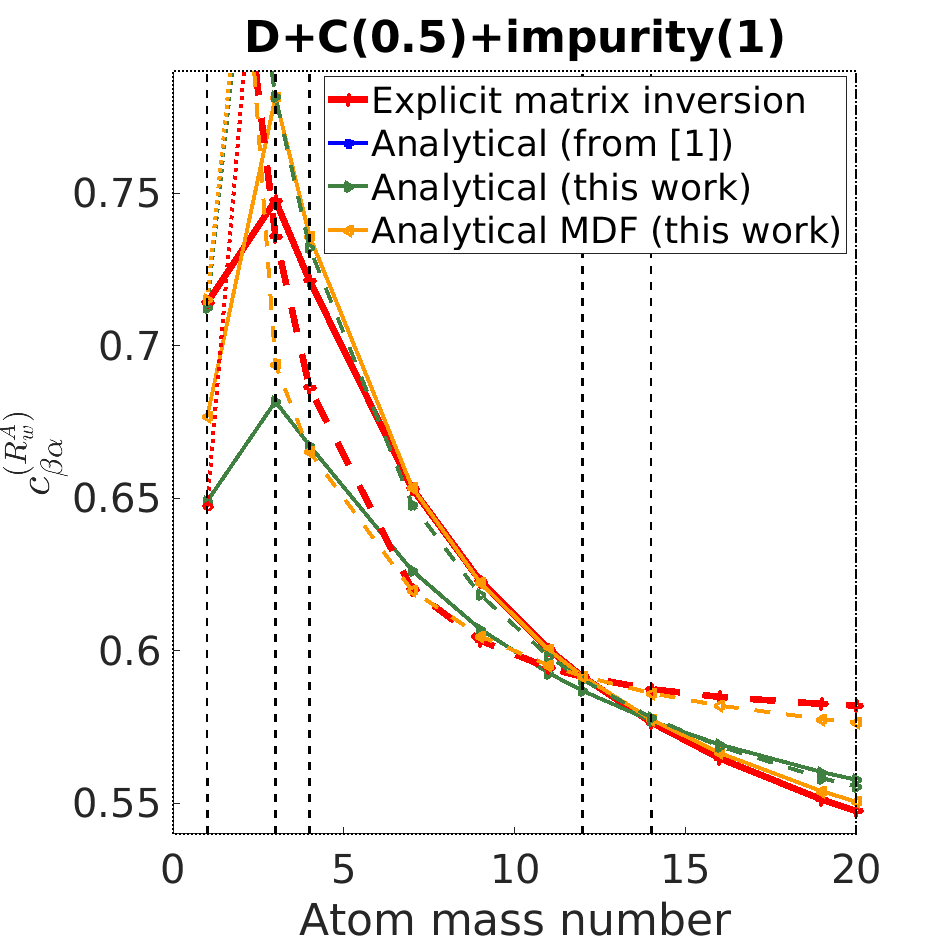} 
\caption{Zoom in the dotted box of Figure~\ref{fig:c_1_D_C_imp}}
\label{fig:c_1_D_C_imp_Zoom_in} 
\end{figure}
The transport coefficient for the friction force between deuterium and impurity predicted by the original Zhdanov analytical expression (8.4.2) \cite{book}, that is used in \cite{Sytova2018,Sytova2020}, is underestimated significantly for all realistic cases compared to the direct numerical solution of the system of equations (8.4.2) in \cite{book} (Figure~\ref{fig:c_1_D_C_imp}): 37\% for helium, 23 \% for carbon, 17\% for neon. This difference in the friction force can affect impurity transport \cite{Senichenkov_2019}.
On the other hand, our improved expression is much closer to the numerical solution (for a discussion of the corresponding deviations see below). This is the result of the inclusion of next order terms in the mass ratio $m_0/m_i$. Indeed, one can see (appendix \ref{app:1}) that, under assumption $m_0/m_i=0$, our improved expression turns into the original Zhdanov formula (8.4.2) in \cite{book}.

Moreover, as it was discussed in the subsection \ref{subsection:heat_flux} (Figure~\ref{fig:c_2_D_C_imp}), for the heavy species, the velocity dependent part of the heat flux is not calculated precisely by our improved expression. The same result can be obtained for the velocity dependent part of the r-moment (5-order vector moment, see Eq. (8.1.2) in \cite{book}). This inaccuracy for the heat flux and r-moment affects the friction force. Therefore, it is necessary to suppress them to zero to get a better match with the numerical result. One way to do this is by setting coefficients \eqref{eq:heat_flux_vel_aver_coef}  and \eqref{eq:r_moment_vel_aver_coef} equal to zero, if $m_\alpha/m_{\beta} <1$.
The result, an even better match (indicated as "Analytical MDF (this work)" in figures~\ref{fig:c_1_D_C_imp} and \ref{fig:c_1_D_C_imp_Zoom_in}, while "Analytical (this work)" corresponds to unmodified heat flux and r-moment) with the numerical result, with the deviation from the matrix inversion solution: 
for tritium 5\% versus 8\%, for helium 2\% versus 8\%, for carbon 0.03\% versus 0.7\%, for neon 0.6\% versus 2\% (for carbon in presence of neon 1\% versus 5\%). However, the mismatch for another impurity/carbon friction force becomes worse (Figure~\ref{fig:c_1_D_C_imp}). Note that the transport coefficient, equal to unity for impurity/impurity interaction (used in \cite{Sytova2018,Sytova2020}), is close to the explicit approach result for most cases. Besides, a correction for the friction force between different charge states can be applied (appendix \ref{app:1}), though for heavy species this transport coefficient is $\approx$0.8-1.0 and significantly differs from unity only for light species.

\subsection{Viscosity}

\begin{figure}
  \centering
\includegraphics[scale=0.3]{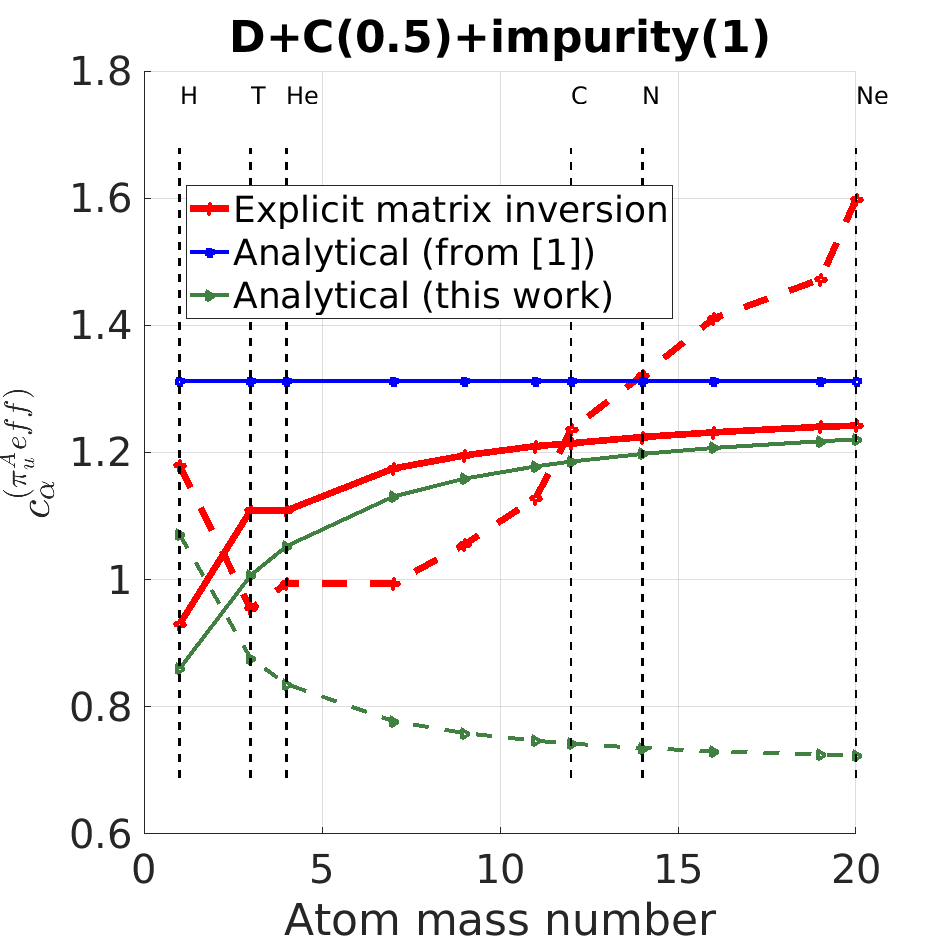}  
\caption{Velocity dependent part of the viscosity transport coefficient for the deuterium (solid) and for another impurity (dashed) using the explicit solution of the system of equations \eqref{eq:aver_viscosity_long_add}-\eqref{eq:aver_sigma_moment_long_add}, the original Zhdanov formula \cite{book}, and our improved formula (appendix \ref{app:1}) for the D + C + another impurity case.
(See the caption to Figure \ref{fig:c_63_D_C_imp} for additional information.)
}
\label{fig:c_pi_1_D_C_imp}  
\end{figure}

\begin{figure}
  \centering
\includegraphics[scale=0.3]{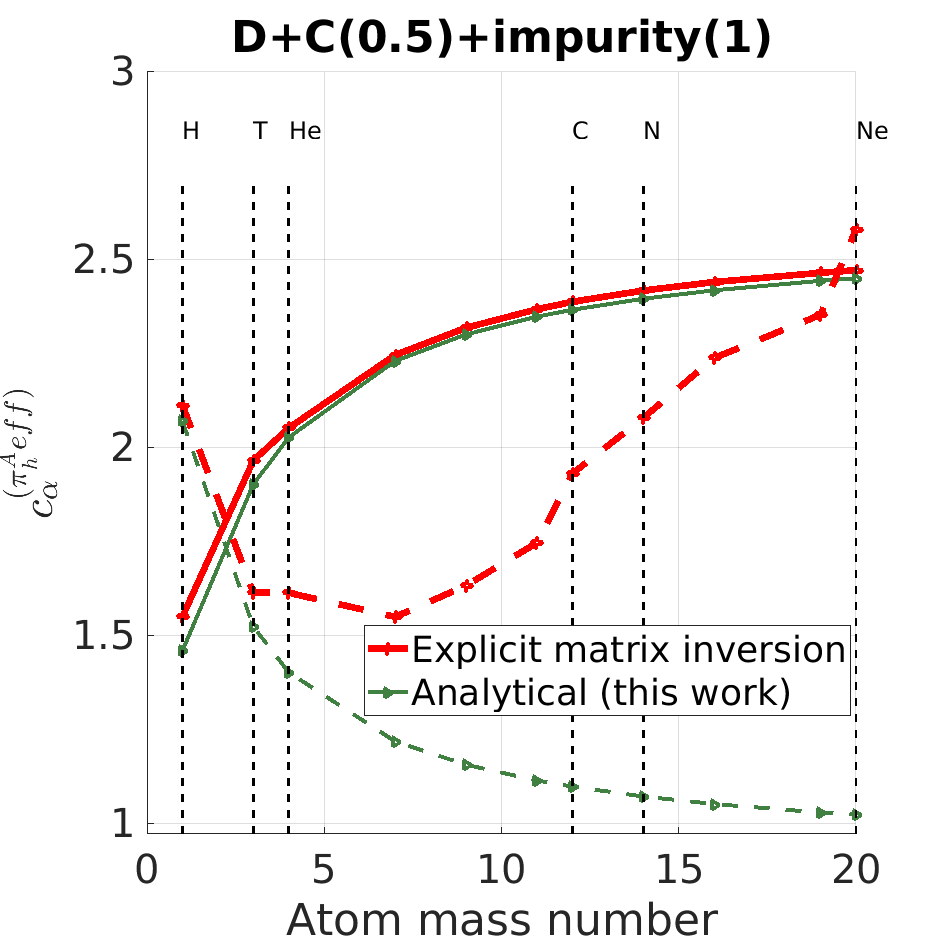}  
\caption{Heat flux dependent part of the viscosity transport coefficient for the deuterium (solid) and for another impurity (dashed) using the explicit solution of the system of equations  \eqref{eq:aver_viscosity_long_add}-\eqref{eq:aver_sigma_moment_long_add} and our improved formula (appendix \ref{app:1}) for the D + C + another impurity case.
(See the caption to Figure \ref{fig:c_63_D_C_imp} for additional information.)
}
\label{fig:c_pi_2_D_C_imp}  
\end{figure}

Finally, using a similar approach for the viscosity equations \eqref{eq:aver_viscosity_long_add} and \eqref{eq:aver_sigma_moment_long_add}, one can obtain analytical expressions for both the velocity and heat flux dependent parts of the viscosity (appendix \ref{app:1}). Now again consider the deuterium, carbon and another impurity case. The viscosity for species $\alpha$ summed over all charge states (for comparison with the analytical result from the matrix inversion method, we assume $W^{\overline h_{\beta}}_{\parallel\parallel}/n_{\beta}=W^{\overline h_{\alpha}}_{\parallel\parallel}/n_{\alpha}$ for arbitrary $\beta$):
\begin{align}
\overline\pi_{\alpha \parallel\parallel}=-p_{\alpha }\tau^{(Zh)}_{\alpha}c^{(\pi_u^Aeff)}_\alpha W_{\parallel\parallel}-\tau^{(Zh)}_{\alpha}c^{(\pi_h^Aeff)}_\alpha W^{\overline h_{\alpha}}_{\parallel\parallel}
\end{align}
In figures~\ref{fig:c_pi_1_D_C_imp} and \ref{fig:c_pi_2_D_C_imp}, $c^{(\pi_u^Aeff)}_\alpha,\ c^{(\pi_h^Aeff)}_\alpha$ are plotted. Other definitions can be found in appendix \ref{app:1}.

It is shown (Figure~\ref{fig:c_pi_1_D_C_imp}) that, for light species, the original Zhdanov expression overestimates the explicit method result significantly, while the new formula provides a closer result. Close answers are obtained for the heat flux dependent part of viscosity using both the improved formula and matrix inversion calculation (Figure~\ref{fig:c_pi_2_D_C_imp}). For this part the analytical expression in \cite{book} was not presented. For the case $m_D/m_{imp}\approx1$ and $n_{imp}/n_D\approx1$, the analytical method works quite well (figures~\ref{fig:c_pi_1_D_C_imp} and \ref{fig:c_pi_2_D_C_imp}). However, for the heavy impurities, the analytical expression
does not match the calculated answer, and this affects the impurity momentum equation. On the other hand, the total momentum equation (summed over all species) is mainly affected by the light component viscosity (in case $m_D/m_{imp}<1$ and $n_{imp}/n_D<1$), and $\mu_{\alpha1}, \mu_{\alpha2}$ as discussed in \cite{helander2005collisional} can be obtained in the Pfirsch-Schlüter regime in the presence of impurities using expressions from appendix \ref{app:1}.

\subsection{Summary}

New analytical expressions discussed in this section and presented in appendix \ref{app:1} can be applied for the cases with one light and several heavy ions, when heat flux and viscosity of the heavy species does not affect the solution (for example $n_{imp}/n_D<1$). For plasmas where the mass of the main component is not significantly different from the mass of the impurity, improved expansions provide solutions closer to the numerical one than can be obtained by the original expression in \cite{book}. Moreover, for the lighter than main ions impurities, the sign of the new thermal force is inverted, therefore our new expressions provide a qualitatively correct result.
Furthermore, new expressions are written in general form without specifying which  species are main ions or impurities, and which species are light or heavy, as it was for original analytical expressions in \cite{book,Sytova2018,Sytova2020}. Besides, these formulae describe transport for each charge state of all the ions. Therefore, it allows us to consider helium plasmas with impurities, which was not possible in the previous SOLPS-ITER model, where main ions were assumed to be singly ionized.
However, either in the case of accurate transport calculation for mixtures with close masses or in the case where high order moments of the heavy species distribution function (heat flux, r-moment (5-order vector moment, see Eq. (8.1.2) in \cite{book}), viscosity, $\sigma$-moment (4-order tensor moment see Eq.\ (8.1.2) in \cite{book})) are important, the explicit approach discussed in the previous section should be applied. 
Finally, it is important to mention that these results can be applied only for the collisional plasmas, since time derivatives and gradients (which are of the order of $(\lambda /L)^2$, where $\lambda$ is a mean free path and $L$ is a space scale of the gradients) were neglected in high order moments equations (heat flux, r-moment, viscosity, $\sigma$-moment) \cite{book}. The only exception, the heat flux gradients are added into the viscosity and $\sigma$-moment equations \eqref{eq:aver_viscosity_long_add}-\eqref{eq:aver_sigma_moment_long_add} for taking into account the Pfirsch-Schlüter regime effects discussed in \cite{helander2005collisional,Hirshman_1981,Hinton1976}. In case of the less collisional plasmas next order terms have to be considered, as well.

\section{Application of Grad's closure to SOLPS-ITER}\label{For_SOlPS}

\subsection{Heat flux, friction and heat exchange terms}

\begin{figure}
  \centering
\includegraphics[scale=0.3]{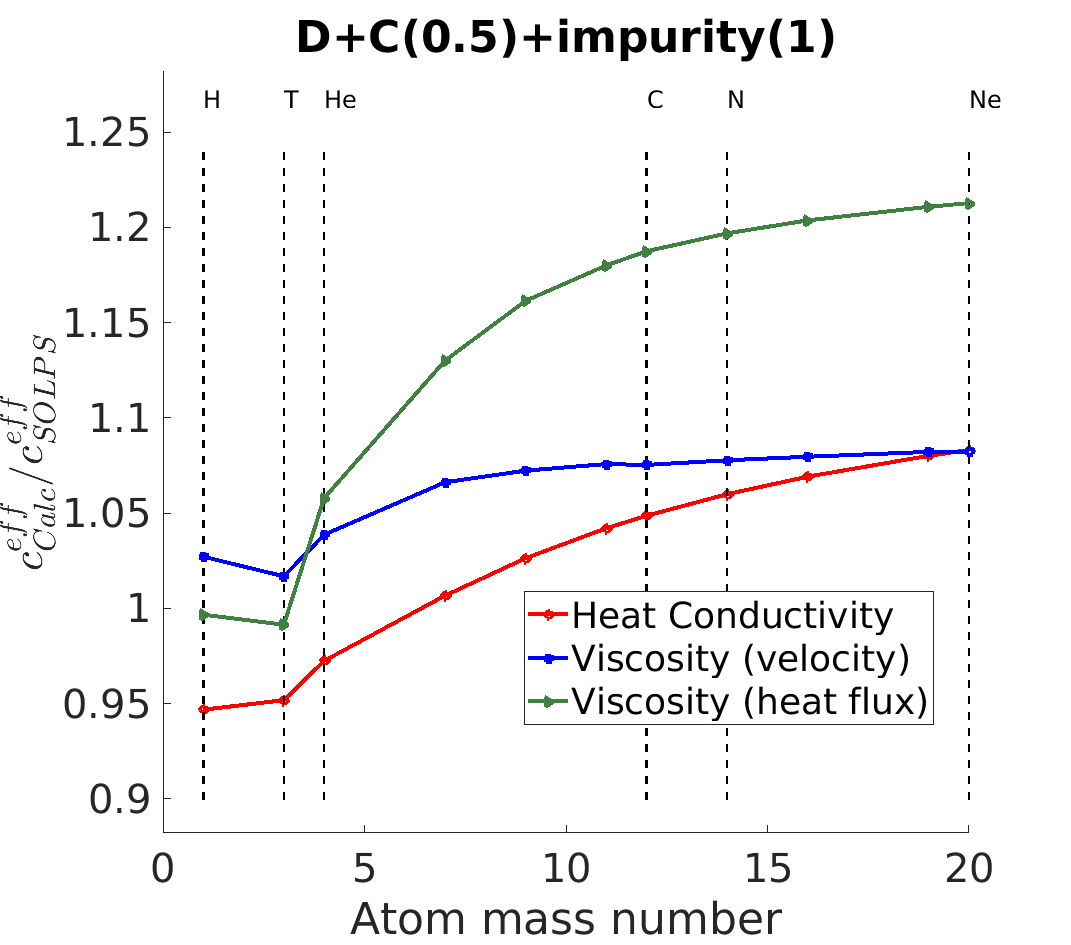}  
\caption{Ratio of the effective transport coefficient calculated using the explicit solution of the system of equations  (8.4.2) in \cite{book} and \eqref{eq:aver_viscosity_long_add}-\eqref{eq:aver_sigma_moment_long_add}  to the current SOLPS-ITER model \cite{SOLPSmanual} for the D + C + another impurity.
Here $h^T_{ions \parallel Calc/SOLPS}=-c^{eff}_{Calc/SOLPS}\nabla_\parallel T^{(Br)}_{ions}$; $\pi_{ions \parallel\parallel Calc/SOLPS}^u=-c^{eff}_{Calc/SOLPS}W_{\parallel\parallel}$; $\pi_{ions\parallel\parallel Calc/SOLPS}^h=-c^{eff}_{Calc/SOLPS}W^{h_{ions}}_{\parallel\parallel}$ (W-tensors see Eq. \eqref{eq:stress_tensor} of appendix \ref{app:1}), where $c^{eff}_{Calc/SOLPS}$ coefficients are chosen using either current SOLPS-ITER expressions or explicit approach of solving system algebraic equations of high order moments (8.4.2) in \cite{book},\eqref{eq:aver_viscosity_long_add}-\eqref{eq:aver_sigma_moment_long_add}
(See the caption to Figure \ref{fig:c_63_D_C_imp} for additional information.)}
\label{fig:c_eff_D_C_imp}  
\end{figure}

First of all, it is important to mention that, in the current SOLPS-ITER model, ion temperatures $T^{(Br)}_{ions}$ are considered to be equal for all ions  \cite{Rozhansky_2009}, therefore \eqref{eq:ion_heat_eq} has to be summed over all ions.  It should be noted that even in this case, $T_{\alpha Z}$ can be different due to \eqref{eq:temp_diff}.
Second, the part of the heat flux that depends on the velocity difference between ions with different masses (last term in \eqref{eq:heat_flux_expression_par}), is not presented in models which are based on the Braginskii equations (where such a term appears only in the electron heat flux) and has been neglected in multicomponent cases \cite{WIESEN2015480,KUKUSHKIN20112865,BUFFERAND2017852,Feng2004,VIOLA2017786}; in future this contribution should be added. Moreover, the effective ion conductivity also changes in plasmas with non-trace-impurities, even under the assumption of identical temperature for all ions.
In Figure~\ref{fig:c_eff_D_C_imp}, it is shown how the ion heat conductivity calculated using the explicit matrix inversion approach be different from what is predicted by the current SOLPS-ITER model (section B.4) \cite{SOLPSmanual}.

A comparison of the friction and thermal force between the current SOLPS-ITER model \cite{Sytova2018,Sytova2020}, our improved expressions, and the explicit approach is made in detail in section \ref{Friction_sec}. Therefore, it is not repeated here. The improved formulae for the friction and thermal forces have been implemented into the SOLPS-ITER code. Test results can be found in section \ref{Test_sec}.

In the current SOLPS-ITER  model (sections C.3.1 and C.4.1) in \cite{SOLPSmanual}, the heat source,  due  to the friction between different ions and with electrons, in heat exchange terms for ions $Q^{(Br)}_{ions}$ and electrons $Q^{(Br)}_{e}$ is written so as to ensure the global conservative property in collisions \cite{braginskii1965transport}. Using \eqref{eq:Q_diff} for electrons and ions, the distribution between electron and ion channels can be found more accurately. One can use \eqref{eq:Q_diff} to find the correct heat distribution between different ions, as well.

\subsection{Viscosity}
Consider an orthogonal curvilinear coordinate system where x is a poloidal coordinate and y is a radial coordinate \eqref{eq:def_geometry}, as it is done in \cite{Rozhansky_2001}.
\begin{align}\label{eq:def_geometry}
b_x=\frac{B_x}{B},\ \ h_x=\frac{1}{||\nabla x|| },\ \ h_y=\frac{1}{||\nabla y|| },\ \ h_z=\frac{1}{||\nabla z|| },\ \  \sqrt{g}=h_xh_yh_z
\end{align}
In this coordinate system \eqref{eq:def_geometry} and under assumption that $n_{\alpha Z}, T_{\alpha Z}, \varphi$ are flux surface functions inside the separatrix where parallel viscosity plays its larger role \cite{helander2005collisional,Hirshman_1981} in the parallel momentum balance, Eq. \eqref{eq:div_viscosity_par} in appendix \ref{app:2} turns into:
\begin{multline}\label{eq:div_viscosity_par_SOLPS}
 (\nabla \cdot\overleftrightarrow{\pi}^{(par)}_{\alpha Z})_\parallel=-\frac{4}{3}b_xB^{3/2}\frac{\partial}{h_x\partial x}\Bigg[\frac{b_x}{B^{2}}\frac{n_{\alpha Z}}{n_{\alpha}}\bigg[\sum_{\beta}\tilde c^{(\pi_u^A)}_{\alpha \beta}\tau^{(Zh)}_{\beta \alpha}p_{\beta}\frac{\partial}{h_x\partial x}\bigg(\frac{\sqrt{B}}{b_x}u_x\bigg)+\\
2c^{(\pi_u^B)}_{\alpha}\tau^{(Zh)}_{\alpha \alpha}\frac{\overline{Z_\alpha^2}}{Z^2}p_{\alpha}\frac{\partial}{h_x\partial x}\bigg(\frac{\sqrt{B}}{b_x}u_x\bigg)\bigg]\Bigg]-\\
\frac{8}{15}b_xB^{3/2}\frac{\partial}{h_x\partial x}\Bigg[\frac{b_x}{B^{2}}\frac{n_{\alpha Z}}{n_{\alpha}}\bigg[\sum_{\beta}\left(\tilde c^{(\pi_h^A)}_{\alpha \beta}\tau^{(Zh)}_{\beta \alpha}\frac{\partial}{h_x\partial x}\bigg(\frac{\sqrt{B}}{b_x}\overline h_{\beta x}\bigg)\right)+\\
2c^{(\pi_h^B)}_{\alpha}\tau^{(Zh)}_{\alpha \alpha}\frac{\overline{Z_\alpha^2}n_\alpha}{Z^2n_{\alpha Z}}\frac{\partial}{h_x\partial x}\bigg(\frac{\sqrt{B}}{b_x}h_{\alpha Z x}\bigg)\bigg]\Bigg]
\end{multline}
Note that expression \eqref{eq:div_viscosity_par_SOLPS} is a generalization of the corresponding terms in the momentum equation in \cite{Rozhansky_2001, Rozhansky_2009} for the multicomponent case.
The new result are compared with that currently applied for SOLPS-ITER in Pfirsch-Schlüter regime \cite{Rozhansky_2009} in Figure~\ref{fig:c_eff_D_C_imp}. Despite the fact that the effect on the viscosity (velocity part) $\pi_{ion}^u$ summed over all ions is not more than 8\% due to the new approach, the effect on the impurity viscosity  (velocity part) is significant. For the high mass high-Z impurity (carbon and higher), the viscosity (velocity part) transport coefficient is $\approx$1.5-2.5 times higher than currently implemented in SOLPS-ITER.
The heat flux dependent part of the viscosity $\pi_{ion}^h$ (whose importance was discussed above) can be 20\% higher for this closure (assuming for comparison $W^{\overline h_{\beta}}_{\parallel\parallel}/n_{\beta}=W^{\overline h_{ion}}_{\parallel\parallel}/n$ for either $\beta$) than for the current SOLPS-ITER approach (Figure~\ref{fig:c_eff_D_C_imp}). The impurity viscosity (heat flux dependent part) can be up to 1.9 times higher. 

\subsection{Test cases using improved expressions}\label{Test_sec}

\begin{figure}

    \begin{subfigure}{0.3\textwidth}
        \includegraphics[scale=0.35]{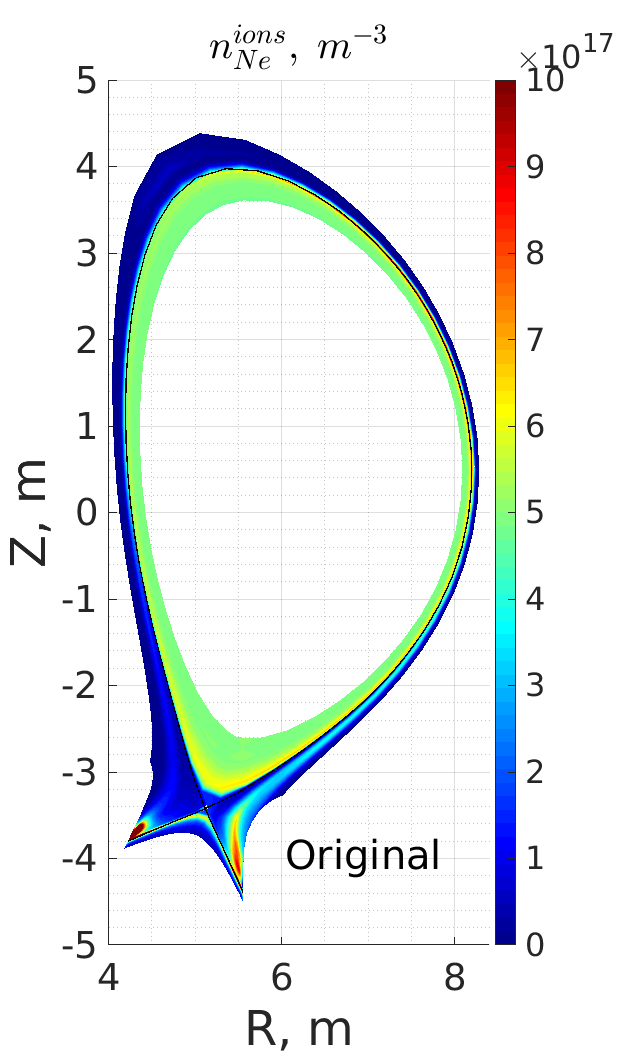}  
        \caption{}
        \label{fig:Iter_nNe_Original}
    \end{subfigure}
 \qquad \qquad \qquad \qquad
    \begin{subfigure}{0.3\textwidth}
        \includegraphics[scale=0.35]{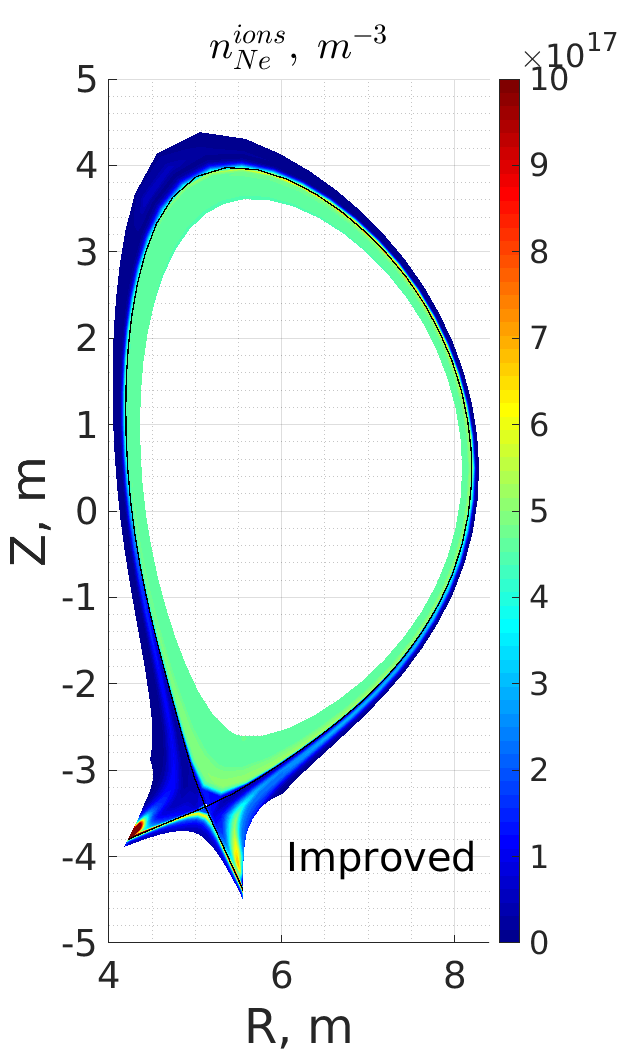}  
        \caption{}
        \label{fig:Iter_nNe_Imrpoved}
    \end{subfigure}
    \caption{2D neon density distribution for the two identical ITER modeling cases using the SOLPS-ITER code, in which the original expressions from \cite{Sytova2018,Sytova2020} are applied (\#123077) (a) and our improved expressions are applied (\#123078) (b)}\label{fig:Ne_dens}.
\end{figure}

In Figure \ref{fig:c_1_D_C_imp_Zoom_in} one can see that, even for deuterium plasmas with neon non-trace-impurity transport coefficient, the friction force calculated by the original Zhdanov formula Eq. (8.4.2) in  \cite{book} is different compared to our improved treatment, while the thermal force remains the same (Figure \ref{fig:Omega_D_C_i_Zeff_corr}), that should affect neon transport in the tokamak. The new friction and thermal force treatments were implemented into the SOLPS-ITER code and tested for a deuterium and neon ITER case. We present two cases, a reference case, run with the original model for the friction and thermal force terms, and a simulation where our new treatment is applied. These cases are cataloged in the ITER Integrated Modelling Analysis Suite (IMAS) database as \#123077 and \#123078, respectively. 
For the test, an intermediate case between cases 1b (\#123014) and 2b (\#123018) (with drifts) was chosen from those that are discussed in \cite{Kaveeva_2020}, with parameters: divertor neutral pressure $p_n=7.5Pa$ and relative Ne concentration, separatrix-averaged, $c_{Ne}=1.0\%$. 
To illustrate the changes in the impurity transport due to the new thermal and friction force treatments, the neon density is plotted in Figure \ref{fig:Ne_dens}.

One can see that, even for the deuterium and neon plasma, the mass difference is not high enough to assume $m_D/m_{Ne}\approx 0$. The application of the new formulas can observably affect the impurity transport for such cases. Also, after implementation of the new expressions, the relative separatrix-averaged Ne concentration dropped to $c_{Ne}=0.8\%$, while the neon seeding rate is kept constant ($6.0\cdot 10^{19}\ particles/s$). In these tests, neon throughput and pumping speed below the Dome are kept constant. However, actual extraction of neon from the divertor by the pumping system, which depends on the distribution between neon flows in the divertor (part goes into the pumping system and part goes upstream), slightly increases, that was seen on the run diagnostics during the simulation. This leads to 10\% drop of the amount of neon in the whole computational domain and in the core in particular. It is interesting to note that the amount of neon in the outer divertor region decreases by 20\%, while the amount of neon in the inner divertor region does not change significantly. This impact on the neon distribution should be investigated.

It is also interesting to note that in the regions with low amounts of neon, the thermal force coefficients are slightly smaller, while the friction force remains the same according to \eqref{eq:tran_coef_trace}, which is different compared to the non-trace-impurity regions, where transport coefficients follow behaviors described in figures \ref{fig:Omega_D_C_i_Zeff_corr} and  \ref{fig:c_1_D_C_imp_Zoom_in}. This example again shows that an accurate treatment of the main component distribution function shape affects the impurity transport both for the trace-impurity and non-trace-impurity cases. We demonstrate these runs to show that our improved expressions for the transport coefficients affect impurity transport in tokamaks like ITER. Further detailed analysis is required to understand this impurity behaviour and is supposed to be conducted in the following papers.

It is necessary to mention that, for this test, the expressions for thermal \eqref{eq:chst_Thermal_force} and friction forces \eqref{eq:chst_Friction_force} in appendix \ref{app:1} were slightly reduced. The thermal force for the impurity type "I" and charge state "Z" is assumed to be:
\begin{align} 
R^T_{I Z}= {c}^{(R_T^A)}_{0I}\frac{Z^2n_{I Z}}{\overline{Z_0^2}}\nabla   T^{(Br)}_{ions}
\end{align}
The friction force for the impurity type "I" and charge state "Z" is assumed as:
\begin{multline}
R_{I Z}^{\textbf w}=
-n_I I_{I Z}\frac{\mu_{I 0}}{\tau^{(Zh)}_{I 0}}(1+{c}^{(R_w^A1)}_{I 0})(\textbf w_{I Z}-{\textbf w}_{0 })-\\
n_I I_{I Z}\sum_{\beta\neq0}\frac{\mu_{I \beta}}{\tau^{(Zh)}_{I \beta}}(\textbf w_{I Z}-\overline{\textbf w}_{\beta })
\end{multline}
where "0" is connected to the main light single charge state ion and thermal and friction forces for the main ions are:
\begin{align}
R^{T}_{0}=
-\sum_{\beta\neq0}n_{0}{c}^{(R_T^A)}_{0\beta}\frac{\overline{Z_\beta^2}n_\beta}{\overline{Z_0^2}n_0}\nabla T^{(Br)}_{ions}
\end{align}
\begin{align}
{R^{\textbf w}_{0}}=-n_0\sum_{\beta\neq0}\frac{\mu_{0 \beta}}{\tau^{(Zh)}_{0 \beta}}(1+{c}^{(R_w^A1)}_{I 0})({{\textbf w}}_{0}-\overline{\textbf w}_{\beta })
\end{align}
On the one hand, such approach allows us to change only transport coefficients for this first test without rewriting equations significantly. On the other hand, for the deuterium and neon case these expressions provide close results to \eqref{eq:chst_Thermal_force} and \eqref{eq:chst_Friction_force} in appendix \ref{app:1}. However, full \eqref{eq:chst_Thermal_force} and \eqref{eq:chst_Friction_force} will be implemented into SOLPS-ITER code later without specifying main ions and impurities explicitly.

\section{Conclusions}

This paper tackles the closure in the parallel direction of the system of fluid equations using Grad's method \cite{grad1963asymptotic} for multicomponent collisional plasmas and  extends the study made in \cite{book}. The method allows one to obtain transport coefficients for the heat flux, friction and viscosity terms for each charge state of each species for the arbitrary plasma composition case, including deuterium + tritium + helium + other impurities mixtures, where densities and masses of species can be comparable. Therefore, this approach can be implemented into SOLPS-ITER \cite{WIESEN2015480,Bonnin2016} or other fluid codes to model this mixture for future reactors (ITER \cite{Ikeda_2007}, DEMO \cite{FEDERICI2014882}, CFETR \cite{Wan2017}) and complex compositions in existing machines.

In contrast to previous papers devoted to such an approach \cite{BUFFERAND201982, bergmann1996implementation}, two major improvements are made: corrections for the direct implementation of the closure \cite{book} into the Braginskii equations; and the heat flux dependent part of viscosity. The importance of this part was discussed in \cite{helander2005collisional,Hirshman_1981}. In this paper the result in collisional regime is extended to arbitrary plasma mixtures.

New analytical expressions were developed (see appendix \ref{app:1}) which provide a better match to the explicit matrix inversion method for the heat flux, friction and viscosity term and extend the applicability of the analytical approach to lower mass impurities. Moreover, this accurate treatment for multi charge state ions allows for the use of our analytical approach for cases where all ions have multiple charge states, for example helium plasmas with impurities, which was not possible with the original expressions.

These new friction and thermal force terms descriptions have been implemented into the SOLPS-ITER code and tested for an ITER deuterium plasma with non-trace-impurity neon and compared with the current SOLPS-ITER model \cite{Sytova2018,Sytova2020}. Even for such a mixture, the new formulae show different impurity transport behavior confirming that the $m_D/m_{Ne}\ll 1$ assumption is not accurate enough. Thus, further studies of the impurity transport in ITER and other devices using this improved approach is required.

\section*{Acknowledgments}

This work has been carried out within the framework of the EUROfusion Consortium and has received funding from the Euratom research and training programme 2014-2018 and 2019-2020 under grant agreement No 633053. This work was performed in part under the auspices of the ITER Scientist Fellow Network. The views and opinions expressed herein do not necessarily reflect those of the European Commission or of the ITER Organization.

\appendix

\section{Appendix}
\label{app:1}
\subsection{Definitions}
Here are defined the variables used in this paper:

Reduced mass:
\begin{align}
\mu_{\alpha \beta}=\frac{m_{\alpha}m_\beta}{m_\alpha+m_\beta}
\end{align}

Averaged over charge states variables:
\begin{align}\label{eq:Z2_def}
\overline{Z_\alpha^2}n_{\alpha}=\sum_ZZ^2n_{\alpha Z},\ \ \ \ I_{\alpha Z}=\frac{Z^2n_{\alpha Z}}{\overline{Z_\alpha^2}n_{\alpha}}
\end{align}
\begin{align}\label{eq:aver_moments}
\overline{\textbf w}_{\alpha}=\sum_{Z}I_{\alpha Z}\textbf w
_{\alpha Z},\ \ \ \ 
\overline{\textbf h}_{\alpha}=\sum_{Z}\frac{p_\alpha I_{\alpha Z}\textbf h_{\alpha Z}}{p_{\alpha Z}}\\
\overline{\textbf r}_{\alpha}=\sum_{Z}\frac{p_\alpha I_{\alpha Z}\textbf r_{\alpha Z}}{p_{\alpha Z}},\ \ \ \ 
\overline \pi_{\alpha  \parallel\parallel}=\sum_{Z}\frac{p_\alpha I_{\alpha Z}\pi_{\alpha Z \parallel\parallel}}{p_{\alpha Z}}
\end{align}
Definitions connected to the moments of the distribution function:
\begin{align}\label{eq:dens_def0}
n_{\alpha Z}=\iiint f_{\alpha Z }d^3\textbf {v},\ \ \ n_{\alpha Z}u_{\alpha Z k}=\iiint v_kf_{\alpha Z }d^3\textbf {v}
\end{align}
\begin{align}\label{eq:aver_vel}
\textbf w_{\alpha Z}=\textbf u_{\alpha Z}-\textbf u
,\ \ \ \ 
\textbf u=\frac{\sum_{\alpha,Z}m_{\alpha}n_{\alpha Z}\textbf u_{\alpha Z }}{\sum_{\alpha,Z}m_{\alpha}n_{\alpha Z}}
\end{align}
\begin{align}\label{eq:dens_def}
n=\sum_{\alpha}n_{\alpha},\ \ n_{\alpha}=\sum_Zn_{\alpha Z},\ \ p_{\alpha Z}=n_{\alpha Z}T_{\alpha Z}
\end{align}
\begin{align}\label{eq:temp_def}
nT=\sum_{\alpha}n_{\alpha}T_{\alpha},\ \ p_{\alpha}=n_{\alpha}T_{\alpha}=\sum_Zn_{\alpha Z}T_{\alpha Z}
\end{align}
\begin{align}
n_{\alpha}\widetilde{\nabla T_{\alpha}}=\sum_Zn_{\alpha Z}\nabla T_{\alpha Z}
\end{align}
Higher order moments are defined in the main text \eqref{eq:def_Zhd_temp}-\eqref{eq:def_Zhd_vis}. $\textbf r_{\alpha Z}$ (5-order vector moment) and $\overleftrightarrow{\sigma}_{\alpha Z}$ (4-order tensor moment) are defined in \cite{book} Eq. (8.1.2).

W-tensors in arbitrary Cartesian coordinate system are: 
\begin{align}\label{eq:stress_tensor}
W_{\parallel\parallel}=B_kB_lW_{kl}/B^2,\ \ \ \ 
W_{kl}=2\bigg[\frac{1}{2}\Big(\frac{\partial u_{k}}{\partial x_l}+\frac{\partial u_{l}}{\partial x_k}\Big)-\frac{1}{3}\delta_{kl}\nabla\cdot \textbf u\bigg]
\end{align}
\begin{align}\label{eq:stress_tensor_h}
W^{\overline h_{\alpha}}_{\parallel\parallel}=B_kB_lW^{\overline h_{\alpha}}_{kl}/B^2,\ \ \ \ 
W^{\overline h_{\alpha}}_{kl}=\frac{4}{5}\bigg[\frac{1}{2}\Big(\frac{\partial\overline h_{\alpha k}}{\partial x_l}+\frac{\partial \overline h_{\alpha l}}{\partial x_k}\Big)-\frac{1}{3}\delta_{kl}\nabla\cdot \overline{ \textbf{h}}_{\alpha}\bigg]
\end{align}
where parallel velocity and drift contributions - diamagnetic and ExB drift velocities - should be taken into account in \eqref{eq:stress_tensor}. Heat flux in \eqref{eq:stress_tensor_h} 
should contain the parallel contribution determined in this paper and the diamagnetic contribution (see appendix \ref{app:2}).

Colllisional right-hand sides of Eq. (8.1.6),(8.1.6') in \cite{book} summed over charge states $\overline{R}^{2*}_{\alpha\parallel\parallel}$ are:
\begin{multline}\label{eq:rhs_viscosity_long_add}
\overline{R}^{20}_{\alpha\parallel\parallel}=\sum_{\beta}\frac{T}{m_\alpha+m_\beta}\Bigg[\frac{\overline G^{(3)}_{\alpha  \beta }\overline\pi_{\alpha \parallel\parallel}}{p_{\alpha }}+\frac{\overline G^{(4)}_{\alpha  \beta }\overline\pi_{\beta  \parallel\parallel}}{p_{\beta }}\\
+\frac{\mu_{\alpha \beta}}{T}\left(\frac{\overline G^{(13)}_{\alpha  \beta }\overline\sigma_{\alpha \parallel\parallel}}{p_{\alpha }}+\frac{\overline G^{(14)}_{\alpha  \beta }\overline\sigma_{\beta  \parallel\parallel}}{p_{\beta }}\right)\Bigg]
\end{multline}
\begin{multline}\label{eq:rhs_sigma_moment_long_add}
\overline{R}^{21}_{\alpha\parallel\parallel}=\sum_{\beta}\frac{T}{m_\alpha+m_\beta}\Bigg[\frac{7}{2}T\mu_{\alpha \beta}\left(\frac{\overline G^{(13)}_{\alpha  \beta }\overline\pi_{\alpha \parallel\parallel}}{m^2_\alpha p_{\alpha }}+\frac{\overline G^{(14)}_{\alpha  \beta }\overline\pi_{\beta \parallel\parallel}}{m^2_\beta p_{\beta }}\right)\\
+\frac{\overline G^{(15)}_{\alpha  \beta }\overline\sigma_{\alpha \parallel\parallel}}{p_{\alpha }}+\frac{\overline G^{(16)}_{\alpha  \beta }\overline \sigma_{\beta  \parallel\parallel}}{p_{\beta }}\Bigg]
\end{multline}
For the G-matrices definitions, see in \cite{book} and corrections in appendix \ref{app:3}.

Collision times are defined by:
\begin{align}
\lambda_{\alpha \beta }=\frac{1}{3}(2\pi)^{-3/2}\overline{Z_\alpha^2}n_\alpha\overline{Z_\beta^2}n_\beta\sqrt{\mu_{\alpha\beta}}\frac{\ln \Lambda}{T^{3/2}}\left(\frac{e^2}{\varepsilon_0}\right)^2,\ \ \ \ 
\tau_{\alpha \beta }^{(Zh)}=\frac{n_\alpha \mu_{\alpha \beta}}{\lambda_{\alpha \beta }}
\end{align}
\begin{align}
\tau_{\alpha Z\beta \zeta}^{(Zh)}=\frac{\tau_{\alpha \beta }^{(Zh)}}{I_{\alpha Z}I_{\beta \zeta}},\ \ \ \tau^{(Zh)}_{\alpha}=\left({\sum_{\beta}\frac{\mu_{\alpha \beta}}{m_\alpha\tau^{(Zh)}_{\alpha \beta}}}\right)^{-1}
\end{align}
Z-variables are defined by:
\begin{align}
Z^*_\alpha=\sum_{\beta \neq \alpha}\frac{\overline{Z_\beta^2}n_\beta}{\overline{Z_\alpha^2}n_\alpha},\ \ \ \ 
Z^*_{c\alpha}=\sum_{\beta \neq \alpha}\sqrt{\frac{\mu_{\alpha\beta}}{m_\alpha}}\frac{\overline{Z_\beta^2}n_\beta}{\overline{Z_\alpha^2}n_\alpha}
\end{align}
\begin{align}
Z^*_{f1\alpha}=Z^s_{2\alpha}=\sum_{\beta\neq\alpha}\left(\frac{\mu_{\alpha \beta}}{m_\alpha}\right)^{3/2}\frac{\overline{Z_\beta^2}n_\beta}{\overline{Z_\alpha^2}n_\alpha}\
\end{align}
\begin{align}
Z^*_{f2\alpha}=Z^s_{8\alpha}=\sum_{\beta\neq\alpha}\left(\frac{\mu_{\alpha \beta}}{m_\alpha}\right)^{5/2}\frac{\overline{Z_\beta^2}n_\beta}{\overline{Z_\alpha^2}n_\alpha}\
\end{align}
\begin{align}
Z^*_{11\alpha}=Z^s_{5\alpha}=\sum_{\beta \neq \alpha}\left(1+\frac{16}{13}\frac{m_\alpha}{m_\beta}+\frac{30}{13}\left(\frac{m_\alpha}{m_\beta}\right)^2\right)\left(\frac{\mu_{\alpha \beta}}{m_\alpha}\right)^{5/2}\frac{\overline{Z_\beta^2}n_\beta}{\overline{Z_\alpha^2}n_\alpha}
\end{align}
\begin{align}
Z^*_{12\alpha}=Z^s_{9\alpha}=\sum_{\beta \neq \alpha}\left(1+\frac{32}{23}\frac{m_\alpha}{m_\beta}+\frac{84}{23}\left(\frac{m_\alpha}{m_\beta}\right)^2\right)\left(\frac{\mu_{\alpha \beta}}{m_\alpha}\right)^{7/2}\frac{\overline{Z_\beta^2}n_\beta}{\overline{Z_\alpha^2}n_\alpha}
\end{align}
\begin{multline}
Z^*_{22\alpha}=Z^s_{11\alpha}=\sum_{\beta \neq \alpha}\bigg(1+\frac{1088}{433}\frac{m_\alpha}{m_\beta}+\frac{3672}{433}\left(\frac{m_\alpha}{m_\beta}\right)^2+\\
\frac{1792}{433}\left(\frac{m_\alpha}{m_\beta}\right)^3+
\frac{1400}{433}\left(\frac{m_\alpha}{m_\beta}\right)^4\bigg)
\left(\frac{\mu_{\alpha \beta}}{m_\alpha}\right)^{9/2}\frac{\overline{Z_\beta^2}n_\beta}{\overline{Z_\alpha^2}n_\alpha}
\end{multline}
\begin{align}
Z^\pi_{11\alpha}=\sum_{\beta \neq \alpha}\left(1+\frac{5}{3}\frac{m_\alpha}{m_\beta}\right)\left(\frac{\mu_{\alpha \beta}}{m_\alpha}\right)^{3/2}\frac{\overline{Z_\beta^2}n_\beta}{\overline{Z_\alpha^2}n_\alpha}\\
Z^\pi_{12\alpha}=\sum_{\beta \neq \alpha}\left(1+\frac{7}{3}\frac{m_\alpha}{m_\beta}\right)\left(\frac{\mu_{\alpha \beta}}{m_\alpha}\right)^{5/2}\frac{\overline{Z_\beta^2}n_\beta}{\overline{Z_\alpha^2}n_\alpha}
\end{align}
\begin{multline}
Z^\pi_{22\alpha}=\sum_{\beta \neq \alpha}\Bigg(1+\frac{185}{51}\frac{m_\alpha}{m_\beta}                                                                                                                                                                                                                                                         
+\frac{154}{51}\left(\frac{m_\alpha}{m_\beta}\right)^2+
\frac{140}{51}\left(\frac{m_\alpha}{m_\beta}\right)^3\Bigg)\left(\frac{\mu_{\alpha \beta}}{m_\alpha}\right)^{7/2}\frac{\overline{Z_\beta^2}n_\beta}{\overline{Z_\alpha^2}n_\alpha}
\end{multline}
Note: If $m_\alpha/m_\beta \rightarrow 0$,  all Z-variables become equal to $Z^*_{\alpha}$ 
\subsection{Results}
Results in this section are presented for the parallel (with regard to the B-field) component (for all components in case $B=0$)
\subsubsection{Heat flux}
Averaged over all charge states:
\begin{align}
\overline{\textbf h}_{\alpha }=-\frac{p_{\alpha }n_{\alpha}}{\lambda_{\alpha\alpha}}c^{(h_T^A)}_{\alpha}\widetilde{\nabla   T_{\alpha}}+
p_{\alpha }\sum_{\beta}c^{(h_w^A)}_{\beta \alpha}(\overline{\textbf w}_{\alpha}-\overline{\textbf w}_{\beta })
\end{align}
where:
\begin{align}\label{eq:heat_flux_vel_aver_coef}
c^{(h_w^A)}_{\beta \alpha}=\frac{25}{16}\frac{\mu_{\alpha \beta}}{m_\alpha}\frac{1}{\tilde{\Delta}_\alpha}\sqrt{2}\sqrt{\frac{\mu_{\alpha\beta}}{m_\alpha}}\frac{\overline{Z_\beta^2}n_\beta}{\overline{Z_\alpha^2}n_\alpha}\Big(\frac{3}{2}-\frac{1}{2}\frac{\mu_{\alpha \beta}}{m_\alpha}+\frac{433\sqrt{2}}{240}Z^*_{22\alpha}-\frac{23\sqrt{2}}{16}\frac{\mu_{\alpha \beta}}{m_\alpha}Z^*_{12\alpha} \Big)
\end{align}
\begin{align}
c^{(h_T^A)}_{\alpha}=\frac{125}{32}\frac{1}{\tilde{\Delta}_\alpha}\bigg(1+\frac{433\sqrt{2}}{360}Z^*_{22\alpha} \bigg)
\end{align}
\begin{align}
\tilde{\Delta}_\alpha=\frac{5629}{1152}Z^*_{11\alpha}Z^*_{22\alpha}-\frac{529}{128}{Z^*_{12\alpha}}^2+\frac{65\sqrt{2}}{32}Z^*_{11\alpha}+\frac{433\sqrt{2}}{288}Z^*_{22\alpha}-\frac{23\sqrt{2}}{16}Z^*_{12\alpha}+1
\end{align}
Note, in the $m_\alpha/m_\beta \rightarrow 0$ limit,  $Z^*_{c\alpha}=Z^*_{11\alpha}=Z^*_{12\alpha}=Z^*_{22\alpha}=Z^*_{\alpha}$ and these expressions turn into (8.4.7) in \cite{book}.

For each charge state:
\begin{align}
 \textbf h_{\alpha Z}=\textbf h_{\alpha Z}^T+\textbf h_{\alpha Z}^{\textbf w}
\end{align}
\begin{align}
\textbf h_{\alpha Z}^T=
-\frac{p_{\alpha Z}n_{\alpha}}{\lambda_{\alpha\alpha}}c^{(h_T^A)}_{\alpha}\widetilde{\nabla   T_{\alpha}}-\frac{p_{\alpha Z}n_{\alpha}}{\lambda_{\alpha\alpha}}c^{(h_T^B)}_\alpha\left(\frac{\overline{Z_\alpha^2}}{Z^2}\nabla T_{\alpha Z}-\widetilde{\nabla T_{\alpha}}\right)
\end{align}
\begin{align}
\textbf h_{\alpha Z}^{\textbf w}=
p_{\alpha Z}\sum_{\beta}c^{(h_w^A)}_{\beta \alpha}(\overline{\textbf w}_{\alpha}-\overline{\textbf w}_{\beta })+p_{\alpha Z}c^{(h_w^B)}_\alpha(\textbf w_{\alpha Z}-\overline{\textbf w}_{\alpha })=
p_{\alpha Z}\sum_{\beta}c^{(h_w)}_{\beta\alpha}(\textbf w_{\alpha Z}-\overline{\textbf w}_{\beta })
\end{align}
where:
\begin{align}
c^{(h_T^B)}_\alpha=-\frac{5}{2}\frac{S^{(11)}_\alpha\lambda_{\alpha\alpha}}{D_\alpha}=\frac{139750}{53471}
\frac{1+\frac{6928\sqrt{2}}{8385}Z^s_{11\alpha}}{D^{part}_\alpha}
\end{align}
\begin{multline}\label{eq:c_6_analyt}
c^{(h_w^B)}_\alpha=\frac{S^{(9)}_{\alpha}S^{(8)}_{\alpha}-S^{(2)}_{\alpha}S^{(11)}_{\alpha}}{D_\alpha}=
\frac{31500}{53471}\\
\frac{1-\frac{139\sqrt{2}}{105}Z^s_{8\alpha}-\frac{46\sqrt{2}}{105}Z^s_{9\alpha}+\frac{1732\sqrt{2}}{1575}Z^s_{11\alpha}+\frac{559\sqrt{2}}{210}Z^s_{2\alpha}-\frac{368}{105}Z^s_{8\alpha}Z^s_{9\alpha}+\frac{6928}{1575}Z^s_{2\alpha}Z^s_{11\alpha}}{D^{part}_\alpha}
\end{multline}
\begin{multline}\label{eq:D_alpha_part}
{D^{part}_\alpha}=\frac{89600}{160413}\frac{D_\alpha}{\lambda_{\alpha\alpha}^2}=\\
1+\frac{204376\sqrt{2}}{160413}Z^s_{11\alpha}+\frac{72670\sqrt{2}}{53471}Z^s_{5\alpha}-\frac{76728\sqrt{2}}{53471}Z^s_{9\alpha}
+\frac{360256}{160413}Z^s_{5\alpha}Z^s_{11\alpha}-\frac{101568}{53471}({Z^s_{9\alpha}})^2
\end{multline}
\begin{align}
c^{(h_w)}_{\beta\alpha}=\delta_{\alpha\beta}c^{(h_w^B)}_{\beta}-\delta_{\alpha\beta}\sum_{\gamma}c^{(h_w^A)}_{\gamma\alpha}+c^{(h_w^A)}_{\beta \alpha}
\end{align}
\subsubsection{R-moment}
The r-moment is a 5-order vector moment, see Eq. (8.1.2) in \cite{book}.

Averaged over all charge states:
\begin{align}
\overline{\textbf r}_{\alpha }=-\frac{p_{\alpha }n_{\alpha}}{\lambda_{\alpha\alpha}}c^{(r_T^A)}_{\alpha}\frac{T}{m_\alpha}\widetilde{\nabla  T_{\alpha}}+
p_{\alpha }\frac{T}{m_\alpha}\sum_{\beta}c^{(r_w^A)}_{\beta \alpha}(\overline{\textbf w}_{\alpha}-\overline{\textbf w}_{\beta })
\end{align}
where:
\begin{align}\label{eq:r_moment_vel_aver_coef}
c^{(r_w^A)}_{\beta \alpha}=-\frac{35}{24}\frac{\mu_{\alpha \beta}}{m_\alpha}\frac{1}{\tilde{\Delta}_\alpha}\sqrt{2}\sqrt{\frac{\mu_{\alpha\beta}}{m_\alpha}}\frac{\overline{Z_\beta^2}n_\beta}{\overline{Z_\alpha^2}n_\alpha}\Big(\frac{5}{2}\frac{\mu_{\alpha \beta}}{m_\alpha}-\frac{3}{2}+\frac{65\sqrt{2}}{16}\frac{\mu_{\alpha \beta}}{m_\alpha}Z^*_{11\alpha}-\frac{69\sqrt{2}}{16}Z^*_{12\alpha} \Big)
\end{align}
\begin{align}
c^{(r_T^A)}_{\alpha}=\frac{175}{48}\frac{1}{\tilde{\Delta}_\alpha}\bigg(1+\frac{23\sqrt{2}}{8}Z^*_{12\alpha} \bigg)
\end{align}
For each charge state:
\begin{align}
 \textbf r_{\alpha Z}=\textbf r_{\alpha Z}^T+\textbf r_{\alpha Z}^{\textbf w}
\end{align}
\begin{multline}
\textbf r_{\alpha Z}^T=
-\frac{p_{\alpha Z}n_{\alpha}}{\lambda_{\alpha\alpha}}c^{(r_T^A)}_{\alpha}\frac{T}{m_\alpha}\widetilde{\nabla  T_{\alpha}}-\frac{p_{\alpha Z}n_{\alpha}}{\lambda_{\alpha\alpha}}c^{(r_T^B)}_\alpha\frac{T}{m_\alpha}\left(\frac{\overline{Z_\alpha^2}}{Z^2}\nabla T_{\alpha Z}-\widetilde{\nabla T_{\alpha}}\right)
\end{multline}
\begin{multline}
\textbf r_{\alpha Z}^{\textbf w}=
p_{\alpha Z}\frac{T}{m_\alpha}\sum_{\beta}c^{(r_w^A)}_{\beta \alpha}(\overline{\textbf w}_{\alpha}-\overline{\textbf w}_{\beta })+p_{\alpha Z}\frac{T}{m_\alpha}c^{(r_w^B)}_\alpha(\textbf w_{\alpha Z}-\overline{\textbf w}_{\alpha })=\\
p_{\alpha Z}\frac{T}{m_\alpha}\sum_{\beta}c^{(r_w)}_{\beta \alpha}(\textbf w_{\alpha Z}-\overline{\textbf w}_{\beta })
\end{multline}
where:
\begin{align}
c^{(r_T^B)}_\alpha=\frac{35}{2}\frac{S^{(9)}_\alpha\lambda_{\alpha\alpha}}{D_\alpha}=\frac{194600}{53471}
\frac{1+\frac{184\sqrt{2}}{139}Z^s_{9\alpha}}{D^{part}_\alpha}
\end{align}
\begin{multline}\label{eq:c_8_analyt}
c^{(r_w^B)}_\alpha=\frac{7S^{(2)}_{\alpha}S^{(9)}_{\alpha}-S^{(5)}_{\alpha}S^{(8)}_{\alpha}}{D_\alpha}=
\frac{17080}{53471}\\
\frac{1+\frac{276\sqrt{2}}{61}Z^s_{9\alpha}+\frac{417\sqrt{2}}{61}Z^s_{2\alpha}-\frac{590\sqrt{2}}{61}Z^s_{8\alpha}-\frac{130\sqrt{2}}{61}Z^s_{5\alpha}+\frac{1104}{61}Z^s_{2\alpha}Z^s_{9\alpha}-\frac{1040}{61}Z^s_{5\alpha}Z^s_{8\alpha}
}{D^{part}_\alpha}
\end{multline}
\begin{align}
c^{(r_w)}_{\beta \alpha}=\delta_{\alpha\beta}c^{(r_w^B)}_{\beta}-\delta_{\alpha\beta}\sum_{\gamma}c^{(r_w^A)}_{\gamma\alpha}+c^{(r_w^A)}_{\beta \alpha}
\end{align}
\subsubsection{Friction term}
Friction terms can be split into thermal force and friction force components:
\begin{align}
\textbf{R}_{\alpha Z}=\textbf{R}_{\alpha Z}^{T}+\textbf{R}_{\alpha Z}^{\textbf w}
\end{align}
Thermal force averaged over all charge states:
\begin{align}\label{eq:fr_term_5}
\textbf{R}_{\alpha}^{T}=
-\sum_{\beta\neq\alpha}\Big[n_{\alpha}c^{(R_T^A)}_{\alpha\beta}\frac{\overline{Z_\beta^2}n_\beta}{\overline{Z_\alpha^2}n_\alpha}\widetilde{\nabla   T_{\alpha}}
-n_{\beta}c^{(R_T^A)}_{\beta\alpha}\frac{\overline{Z_\alpha^2}n_\alpha}{\overline{Z_\beta^2}n_\beta}\widetilde{\nabla T_{\beta}}\Big]
\end{align}
where:
\begin{multline}
c^{(R_T^A)}_{\alpha\beta}=
\frac{1}{\tilde{\Delta}_\alpha}\frac{25\sqrt{2}}{16}\left(\frac{\mu_{\alpha \beta}}{m_{\alpha}}\right)^{3/2}\Big[\frac{3}{2}\bigg(1+\frac{433\sqrt{2}}{360}Z^*_{22\alpha} \bigg)-\frac{1}{2}\frac{\mu_{\alpha \beta}}{m_\alpha}\bigg(1+\frac{23\sqrt{2}}{8}Z^*_{12\alpha} \bigg)\Big]
\end{multline}
If $m_\alpha/m_\beta \rightarrow 0$,  $Z^*_{f1\alpha}=Z^*_{f2\alpha}=Z^*_{c\alpha}=Z^*_{11\alpha}=Z^*_{12\alpha}=Z^*_{22\alpha}=Z^*_{\alpha}$ and:
\begin{align}
\sum_{\beta\neq\alpha}c^{(R_T^A)}_{\alpha\beta}\frac{\overline{Z_\beta^2}n_\beta}{\overline{Z_\alpha^2}n_\alpha}=\frac{25\sqrt{2}}{16}\frac{1}{{\Delta}_\alpha}Z^*_{\alpha}\bigg(1+\frac{11\sqrt{2}}{30}Z^*_{\alpha} \bigg)
\end{align}
which matches the corresponding (8.4.7) in \cite{book}.

For each charge state:
\begin{multline}\label{eq:chst_Thermal_force}
\textbf{R}_{\alpha Z}^{T}=
-n_{\alpha Z}\frac{Z^2}{\overline{Z_\alpha^2}}\sum_{\beta\neq\alpha}\Big[c^{(R_T^A)}_{\alpha\beta}\frac{\overline{Z_\beta^2}n_\beta}{\overline{Z_\alpha^2}n_\alpha}\widetilde{\nabla  T_{\alpha}}
-\frac{n_{\beta}}{n_{\alpha}}c^{(R_T^A)}_{\beta\alpha}\frac{\overline{Z_\alpha^2}n_\alpha}{\overline{Z_\beta^2}n_\beta}\widetilde{\nabla T_{\beta}}\Big]-n_{\alpha Z}c^{(R_T^B)}_{\alpha}\left(\nabla T_{\alpha Z}-\frac{Z^2}{\overline{Z_\alpha^2}}\widetilde{\nabla T_{\alpha}}\right)
\end{multline}
where
\begin{multline}
c^{(R_T^B)}_{\alpha}=\frac{3}{5}c^{(h_T^B)}_\alpha\left(\frac{1
}{2}+\sqrt{2}Z^*_{f1\alpha}\right)- \frac{3}{14}c^{(r_T^B)}_\alpha\left(\frac{1
}{4}+\sqrt{2}Z^*_{f2\alpha}\right)=\\
\frac{31500
}{53471}\frac{\frac{559
}{420}\left(1+\frac{6928\sqrt{2}}{8385}Z^s_{11\alpha}\right)\left(1+2\sqrt{2}Z^*_{f1\alpha}\right)-\frac{139}{420}\left(1+\frac{184\sqrt{2}}{139}Z^s_{9\alpha}\right)\left(1+4\sqrt{2}Z^*_{f2\alpha}\right)}{D^{part}_\alpha}
\end{multline}
Friction force averaged over all charge states:
\begin{align}\label{eq:fric_force_aver}
\textbf{R}^{\textbf w}_{\alpha}=
-n_\alpha\sum_{\beta}\frac{\mu_{\alpha \beta}}{\tau^{(Zh)}_{\alpha \beta}}c^{(R_w^A)}_{\beta \alpha}(\overline{{\textbf w}}_{\alpha}-\overline{\textbf w}_{\beta })
\end{align}
where:
\begin{align}
c^{(R_w^A)}_{\beta \alpha}=1+c^{(R_w^A1)}_{\beta \alpha}
+c^{(R_w^A2)}_{\beta \alpha}+c^{(R_w^A3)}_{\beta \alpha}
\end{align}
\begin{align}
c^{(R_w^A1)}_{\beta \alpha}=
-\sum_{\gamma\neq\alpha}\frac{\mu_{\alpha \gamma}}{m_{\alpha}}\sqrt{\frac{\mu_{\alpha\gamma}}{\mu_{\alpha\beta}}}\frac{\overline{Z_\gamma^2}n_\gamma}{\overline{Z_\beta^2}n_\beta}\Big[\frac{3}{5}
c^{(h_w^A)}_{\beta \alpha}
-\frac{3}{14}\frac{\mu_{\alpha \gamma}}{m_{\alpha}}c^{(r_w^A)}_{\beta \alpha}\Big]
\end{align}
\begin{align}
c^{(R_w^A2)}_{\beta \alpha}=
\sum_{\gamma\neq\alpha}\frac{\mu_{\alpha \gamma}}{m_{\gamma}}\sqrt{\frac{\mu_{\alpha\gamma}}{\mu_{\alpha\beta}}}\frac{\overline{Z_\gamma^2}n_\gamma}{\overline{Z_\beta^2}n_\beta}\Big[\frac{3}{5}
c^{(h_w^A)}_{\beta \gamma}
-\frac{3}{14}\frac{\mu_{\alpha \gamma}}{m_{\gamma}}c^{(r_w^A)}_{\beta \gamma}\Big]
\end{align}
\begin{align}
c^{(R_w^A3)}_{\beta \alpha}=
-\frac{\mu_{\alpha \beta}}{m_{\beta}}\sum_{\gamma}\Big[\frac{3}{5}
c^{(h_w^A)}_{\gamma\beta}
-\frac{3}{14}\frac{\mu_{\alpha \beta}}{m_{\beta}}c^{(r_w^A)}_{\gamma\beta}\Big]
\end{align}
Note:
\begin{multline}
1+c^{(R_w^A1)}_{\beta \alpha}=
1-\frac{15\sqrt{2}}{16}Z^*_{f1\alpha}\frac{\mu_{\alpha \beta}}{m_\alpha}\frac{1}{\tilde{\Delta}_\alpha}\Big(\frac{3}{2}-\frac{1}{2}\frac{\mu_{\alpha \beta}}{m_\alpha}+\frac{433\sqrt{2}}{240}Z^*_{22\alpha}-\frac{23\sqrt{2}}{16}\frac{\mu_{\alpha \beta}}{m_\alpha}Z^*_{12\alpha} \Big)\\
-
\frac{5\sqrt{2}}{16}Z^*_{f2\alpha}\frac{\mu_{\alpha \beta}}{m_\alpha}\frac{1}{\tilde{\Delta}_\alpha}\Big(\frac{5}{2}\frac{\mu_{\alpha \beta}}{m_\alpha}-\frac{3}{2}+\frac{65\sqrt{2}}{16}\frac{\mu_{\alpha \beta}}{m_\alpha}Z^*_{11\alpha}-\frac{69\sqrt{2}}{16}Z^*_{12\alpha} \Big)
\end{multline}
In the $m_\alpha/m_\beta \rightarrow 0$ limit,  $Z^*_{f1\alpha}=Z^*_{f2\alpha}=Z^*_{c\alpha}=Z^*_{11\alpha}=Z^*_{12\alpha}=Z^*_{22\alpha}=Z^*_{\alpha}$ and:
\begin{align}
c^{(R_w^A)}_{\beta \alpha}=1+c^{(R_w^A1)}_{\beta \alpha}={c}^{(1)}_{\beta \alpha}=\frac{1}{{\Delta}_\alpha}\bigg(\frac{2}{9}{Z^*_\alpha}^2+\frac{61\sqrt{2}}{72}Z^*_\alpha+1\bigg)
\end{align}
which matches with (8.4.7) in \cite{book}.

For each charge state:
\begin{multline}\label{eq:chst_Friction_force}
\textbf{R}_{\alpha Z}^{\textbf w}=-n_\alpha I_{\alpha Z}\sum_{\beta}\frac{\mu_{\alpha \beta}}{\tau^{(Zh)}_{\alpha \beta}}c^{(R_w^A)}_{\beta \alpha}(\overline{{\textbf w}}_{\alpha}-\overline{\textbf w}_{\beta })
-n_\alpha I_{\alpha Z}(\textbf w_{\alpha Z}-\overline{\textbf w}_{\alpha })\sum_{\beta}\frac{\mu_{\alpha \beta}}{\tau^{(Zh)}_{\alpha \beta}}
c^{(R_w^B)}_{\beta\alpha}=\\
-n_\alpha I_{\alpha Z}\sum_{\beta}\frac{\mu_{\alpha \beta}}{\tau^{(Zh)}_{\alpha \beta}}c^{(R_w)}_{\beta\alpha}(\textbf w_{\alpha Z}-\overline{\textbf w}_{\beta })
\end{multline}
\begin{align}
c^{(R_w^B)}_{\beta\alpha}=1-\frac{3}{5}\frac{\mu_{\alpha \beta}}{m_{\alpha}}c^{(h_w^B)}_\alpha
+\frac{3}{14}\left(\frac{\mu_{\alpha \beta}}{m_{\alpha}}\right)^2c^{(r_w^B)}_\alpha
\end{align}
\begin{align}
c^{(R_w)}_{\beta\alpha}=
\delta_{\alpha\beta}\sum_{\gamma}\sqrt{\frac{\mu_{\alpha \gamma}}{\mu_{\alpha \beta}}}\frac{\overline{Z_\gamma^2}n_\gamma}{\overline{Z_\beta^2}n_\beta}[c^{(R_w^B)}_{\gamma\alpha}-c^{(R_w^A)}_{\gamma \alpha}]+c^{(R_w^A)}_{\beta \alpha}
\end{align}
\subsubsection{Viscosity}
Averaged over all charge states
\begin{align}
\overline\pi_{\alpha \parallel\parallel}=-\frac{m_\alpha n_{\alpha }}{\lambda_{\alpha\alpha}}c^{(\pi_u^A)}_\alpha p_{\alpha }W_{\parallel\parallel}-\frac{m_\alpha n_{\alpha }}{\lambda_{\alpha\alpha}}c^{(\pi_h^A)}_\alpha W^{\overline h_{a}}_{\parallel\parallel}
\end{align}
where:
\begin{align}
\Delta_\alpha^\pi=\frac{204}{89}Z^\pi_{11\alpha}Z^\pi_{22\alpha}-\frac{108}{89}{Z^\pi_{12\alpha}}^2+\frac{205\sqrt{2}}{178}Z^\pi_{11\alpha}+\frac{102\sqrt{2}}{89}Z^\pi_{22\alpha}-\frac{54\sqrt{2}}{89}Z^\pi_{12\alpha}+1
\end{align}
\begin{align}
c^{(\pi_u^A)}_\alpha=\frac{1025}{1068}\frac{1}{\Delta_\alpha^\pi}\bigg(1+\frac{204\sqrt{2}}{205}Z^\pi_{22\alpha} \bigg)
\end{align}
\begin{align}
c^{(\pi_h^A)}_\alpha=\frac{1655}{1068}\frac{1}{\Delta_\alpha^\pi}\bigg(1+\frac{204\sqrt{2}}{331}Z^\pi_{22\alpha}+\frac{252\sqrt{2}}{331}Z^\pi_{12\alpha} \bigg)
\end{align}
If $m_\alpha/m_\beta \rightarrow 0$,  $Z^*_{c\alpha}=Z^\pi_{11\alpha}=Z^\pi_{12\alpha}=Z^\pi_{22\alpha}=Z^*_{\alpha}$ and:
\begin{align}
(1+\sqrt{2}Z^*_{c\alpha}) c^{(\pi_u^A)}_\alpha=\frac{1025}{1068}\frac{(1+\sqrt{2}Z^*_{\alpha})(1+\frac{204\sqrt{2}}{205}Z^*_{\alpha})}{\frac{96}{89}{Z^*_{\alpha}}^2+\frac{301\sqrt{2}}{178}Z^*_{\alpha}+1}
\end{align}
which matches the last expression on page 181 of \cite{book}.

For each charge state:
\begin{align}
\pi_{\alpha Z \parallel\parallel}=\pi_{\alpha Z \parallel\parallel}^u+\pi_{\alpha Z \parallel\parallel}^h
\end{align}
\begin{align}
\pi_{\alpha Z \parallel\parallel}^u=
-\frac{m_\alpha n_{\alpha Z} n_{\alpha } T}{\lambda_{\alpha\alpha}}\Bigg[c^{(\pi_u^A)}_\alpha+c^{(\pi_u^B)}_{\alpha}\left(\frac{\overline{Z_\alpha^2}}{Z^2}-1\right)\Bigg]W_{\parallel\parallel}
\end{align}
\begin{align}
\pi_{\alpha Z \parallel\parallel}^h=
-\frac{m_\alpha n_{\alpha Z}}{\lambda_{\alpha\alpha}}\left[c^{(\pi_h^A)}_\alpha W^{\overline h_{a}}_{\parallel\parallel}+
c^{(\pi_h^B)}_{\alpha}\Bigg(\frac{\overline{Z_\alpha^2}n_\alpha}{Z^2n_{\alpha Z}}W^{h_{\alpha Z}}_{\parallel\parallel}-W^{\overline h_{\alpha}}_{\parallel\parallel}\Bigg)\right]
\end{align}
where:
\begin{align}\label{eq:c7_part}
c^{(\pi_u^B)}_{\alpha}=-\lambda_{\alpha\alpha}\frac{S^{(15)}_{\alpha}}{{D^\pi_\alpha}}=\frac{265}{334}
\frac{1+\frac{204\sqrt{2}}{265}Z^\pi_{22\alpha}}{D^{\pi part}_\alpha}
\end{align}
\begin{align}\label{eq:c12_part}
c^{(\pi_h^B)}_{\alpha}=-\lambda_{\alpha\alpha}\frac{S^{(15)}_{\alpha}-\frac{7}{2}S^{(13)}_{\alpha}}{{D^\pi_\alpha}}=\frac{475}{334}
\frac{1+\frac{252\sqrt{2}}{475}Z^\pi_{12\alpha}+\frac{204\sqrt{2}}{475}Z^\pi_{22\alpha}}{D^{\pi part}_\alpha}
\end{align}
\begin{multline}\label{eq:D_vis_alpha_part}
{D^{\pi part}_\alpha}=\frac{70}{167}\frac{D^\pi_\alpha}{\lambda_{\alpha\alpha}^2}=\\
1+\frac{159\sqrt{2}}{167}Z^\pi_{11\alpha}+\frac{816\sqrt{2}}{835}Z^\pi_{22\alpha}-\frac{108\sqrt{2}}{167}Z^\pi_{12\alpha}
+\frac{1224}{835}Z^\pi_{11\alpha}Z^\pi_{22\alpha}-\frac{648}{835}(Z^\pi_{12\alpha})^2
\end{multline}
\subsection{S-coefficients}
This section gives coefficients used for results in the previous section.
They are used in \cite{book} and \cite{BUFFERAND201982}. These coefficients are obtained without additional assumptions, therefore they must match exactly the corresponding in \cite{book} and \cite{BUFFERAND201982}.

For the heat flux:
\begin{align}\label{eq:S_2_analyt}
S^{(2)}_\alpha=\sum_{\beta}\frac{5}{2}\frac{\mu_{\alpha \beta}}{m_\alpha} \overline  G^{(2)}_{\alpha \beta}=
\lambda_{\alpha\alpha}
\Big[\frac{3}{4}+\frac{3\sqrt{2}}{2}Z^s_{2\alpha}\Big]
\end{align}
\begin{align}\label{eq:S_5_analyt}
S^{(5)}_\alpha=\sum_{\beta}\overline  G^{(5)}_{\alpha \beta }=
-\lambda_{\alpha\alpha}
\Big[\frac{59}{40}+\frac{13\sqrt{2}}{10}Z^s_{5\alpha}\Big]
\end{align}
\begin{align}\label{eq:S_8_analyt}
S^{(8)}_\alpha=\sum_{\beta}\frac{35}{2}\left(\frac{\mu_{\alpha \beta}}{m_\alpha}\right)^2 \overline {G}^{(8)}_{\alpha \beta }=
-\lambda_{\alpha\alpha}
\Big[\frac{15}{16}+\frac{15\sqrt{2}}{4}Z^s_{8\alpha}\Big]
\end{align}
\begin{align}\label{eq:S_9_analyt}
S^{(9)}_\alpha=\sum_{\beta}\frac{\mu_{\alpha \beta}}{m_\alpha}\overline  G^{(9)}_{\alpha  \beta}=
\lambda_{\alpha\alpha}
\Big[\frac{417}{1120}+\frac{69\sqrt{2}}{140}Z^s_{9\alpha}\Big]
\end{align}
\begin{align}\label{eq:S_11_analyt}
S^{(11)}_\alpha=\sum_{\beta}\overline  G^{(11)}_{\alpha  \beta}=
-\lambda_{\alpha\alpha}
\Big[\frac{1677}{896}+\frac{433\sqrt{2}}{280}Z^s_{11\alpha}\Big]
\end{align}
\begin{align}\label{eq:D_alpha_analyt}
{D_\alpha}=S^{(5)}_{\alpha}S^{(11)}_{\alpha}-7(S^{(9)}_{\alpha})^2
\end{align}
For viscosity:
\begin{align}\label{eq:S_3_analyt}
S^{(3)}_\alpha=\sum_{\beta}\frac{m_\alpha}{m_\alpha+m_\beta}\overline G^{(3)}_{\alpha  \beta}=
-\lambda_{\alpha\alpha}
\Big[\frac{8}{5}+\frac{6\sqrt{2}}{5}Z^\pi_{11\alpha}\Big]
\end{align}
\begin{align}\label{eq:S_13_analyt}
S^{(13)}_\alpha=\sum_{\beta}\frac{\mu_{\alpha \beta}}{m_\alpha+m_\beta}\overline G^{(13)}_{\alpha  \beta}=
\lambda_{\alpha\alpha}
\Big[\frac{3}{7}+\frac{18\sqrt{2}}{35}Z^\pi_{12\alpha}\Big]
\end{align}
\begin{align}\label{eq:S_15_analyt}
S^{(15)}_\alpha=\sum_{\beta}\frac{m_{\alpha}}{m_\alpha+m_\beta}\overline G^{(15)}_{\alpha  \beta}=
-\lambda_{\alpha\alpha}
\Big[\frac{53}{28}+\frac{51\sqrt{2}}{35}Z^\pi_{22\alpha}\Big]
\end{align}
\begin{align}\label{eq:D_vis_alpha_analyt}
{D^\pi_\alpha}={S^{(3)}_{\alpha}S^{(15)}_{\alpha}-\frac{7}{2}\left(S^{(13)}_{\alpha}\right)^2}
\end{align}
\section{Appendix}
\label{app:2}
This appendix considers the study of the divergence $\nabla \cdot \overleftrightarrow{\pi}_{\alpha Z}^{(par)}$, taking into account parallel and drift components of the velocity and parallel and diamagnetic components of the heat flux.

The parallel viscosity \eqref{eq:viscosity_result} in an arbitrary coordinate system is:
\begin{multline}
 \label{eq:vis_par_form}
 {\pi_{\alpha Z}^{(par)}}_{kl} = -\frac32\left(b_kb_l-\frac13\delta_{kl}\right)\bigg[\frac{n_{\alpha Z}}{n_{\alpha}}\left(\sum_{\beta}\big[\tilde{c}^{(\pi_u^A)}_{\alpha\beta}+2c^{(\pi_u^B)}_{\alpha}\frac{\overline{Z_\alpha^2}}{Z^2}\delta_{\alpha\beta}\big]\tau^{(Zh)}_{\beta \alpha}p_{\beta}\right)W_{\parallel\parallel}\\
+\frac{n_{\alpha Z}}{n_{\alpha}}\sum_{\beta}\left(\tilde{c}^{(\pi^A_h)}_{\alpha\beta}\tau^{(Zh)}_{\beta \alpha}W^{\overline h_{\beta}}_{\parallel\parallel}
\right)+2c^{(\pi^B_h)}_{\alpha}\frac{\overline{Z_\alpha^2}}{Z^2}\tau^{(Zh)}_{\alpha \alpha}W^{h_{\alpha Z}}_{\parallel\parallel}\bigg]
\end{multline} 
where $\textbf b=\textbf B/B$ and for $\textbf u$ we substitute into $W_{\parallel\parallel}$ the sum of the parallel velocity, $E\times B$ and
diamagnetic velocities:

\begin{equation}
 \textbf u = \textbf b\cdot \frac{1}{\rho}\sum_{\beta, \zeta} m_\beta n_{\beta\zeta} {u_{\beta\zeta}}_\parallel+
\textbf u^{\rm dia} + 
 \textbf u^{E\times B} 
\end{equation}
\begin{equation}
\textbf u^{\rm dia}=\frac{1}{\rho B^2}\left[\textbf B \times \nabla\sum_{\beta, \zeta}
 \left(\frac{m_\beta n_{\beta\zeta} T_{\beta\zeta}}{\zeta e}\right)\right]
\end{equation}
\begin{equation}
  \textbf u^{E\times B} =\frac{1}{B^2}\left[\textbf B \times \nabla\varphi\right]
\end{equation}
where summation is over all ions $\beta$ with charge states $\zeta$ and

$${u_{\beta\zeta}}_\parallel=\left(\textbf b\cdot
 \textbf u_{\beta\zeta} \right), \qquad \rho = \sum_{\beta, \zeta} m_\beta n_{\beta\zeta}.$$
For $\overline{\textbf h}_{\alpha}$ we substitute into $W^{\overline h_{a}}_{\parallel\parallel}$ (using \eqref{eq:aver_moments}), and for ${\textbf h}_{\alpha Z}$ we substitute into $W^{ h_{\alpha Z}}_{\parallel\parallel}$:

\begin{equation}
{\textbf h}_{\alpha Z}=\textbf b\cdot h_{\alpha Z\parallel}+ {\textbf h}_{\alpha Z\perp}^{(dia)}
\end{equation}
where (see Eq. (8.3.1) of \cite{book}):
\begin{align}
\textbf h^{(dia)}_{\alpha Z \perp}=-\frac{5}{2}\frac{p_{\alpha Z}[\nabla T_{\alpha Z}\times\textbf B]}{ZeB^2}
\end{align}
Making use of $$ \left(\vec{\bf B}\cdot\nabla
\right)\vec{\bf B} = \frac12\nabla B^2 - \left[\vec{\bf B}\times\left[\nabla\times\vec{\bf B}\right]\right] \approx \frac12\nabla B^2,$$
one finally gets:
\begin{multline}\label{eq:div_viscosity_par}
 (\nabla \cdot\overleftrightarrow{\pi}^{(par)}_{\alpha Z})_\parallel= 
 -\frac{2}{3}B^{3/2}\nabla_\parallel\Bigg[
 \frac{n_{\alpha Z}}{n_{\alpha}}
 \bigg[\sum_{\beta}\tilde c^{(\pi_u^A)}_{\alpha \beta}\tau^{(Zh)}_{\beta \alpha}p_{\beta}+
 2c^{(\pi_u^B)}_{\alpha}\tau^{(Zh)}_{\alpha \alpha}\frac{\overline{Z_\alpha^2}}{Z^2}p_{\alpha} \bigg] \ast \\
\left[\frac{1}{B^2}\nabla_\parallel\bigg(2\sqrt{B}u_\parallel\bigg) - 
\left( \left( \textbf u^{E\times B} +
\textbf u^{\rm dia}\right)\cdot\frac{\nabla B}{B^{5/2}}\right) + 
\frac{1}{B^{3/2}}\left(\textbf u^{\rm dia}\cdot\frac{\nabla\rho}{\rho}\right)\right]
\Bigg]-\\ 
-
 \frac{4}{15}B^{3/2}\nabla_\parallel\Bigg[\frac{n_{\alpha Z}}{n_{\alpha}}\bigg[\sum_{\beta}\left(\tilde c^{(\pi_h^A)}_{\alpha \beta}\tau^{(Zh)}_{\beta \alpha}\left(\frac{1}{B^2}\nabla_\parallel\bigg(2\sqrt{B}{\overline h_\beta}_\parallel\bigg)\right)-
 \left(\overline{\textbf h}_{\alpha\perp}^{dia} \cdot \frac{\nabla B}{B^{5/2}}\right) - \right. \\ \left.
 \frac52\sum_\zeta\left(\left[\textbf B\times\nabla T_{\beta \zeta}\right]\cdot\nabla
 \left(\frac{\zeta n_{\beta \zeta}T_{\beta \zeta}}{\overline Z_\beta^2eB^2}\right)\right)\right) - 
 2c^{(\pi_h^B)}_{\alpha}\tau^{(Zh)}_{\alpha \alpha}\frac{\overline{Z_\alpha^2}n_\alpha}{Z^2n_{\alpha Z}}\left(\frac{1}{B^2}\nabla_\parallel\bigg(2\sqrt{B}{h_{\alpha Z}}_\parallel\bigg)+ \right. \\ 
 \left. \left( \textbf h_{\alpha Z \perp}^{dia} \cdot \frac{\nabla B}{B^{5/2}}\right) -
 \frac52\left(\left[\textbf B\times\nabla T_{\alpha Z}\right]\cdot\nabla
 p_{\alpha Z}\right)
 \right)\bigg]\Bigg]
\end{multline}

\section{Appendix}
\label{app:3}

Here we provide corrections to mistakes found in the 8th chapter of the original monograph \cite{book}.

Note, in this appendix \ref{app:3}, the temperature is given in Kelvin following the system of units used in \cite{book}.

In the expressions (8.1.4), the numerator in the second term has to be changed.  The correct numerator is 136, not 139:
\begin{align}
G^{(11)}_{\alpha Z \beta \zeta}=-\left(\frac{433}{280}\frac{m^2_\beta}{m^2_\alpha}+\frac{136}{35}\frac{m_\beta}{m_\alpha}+\frac{459}{35}+\frac{32}{5}\frac{m_\alpha}{m_\beta}+5\frac{m^2_\alpha}{m^2_\beta}\right)\kappa^2_{\alpha\beta}\lambda_{\alpha Z \beta \zeta}
\end{align}
R.h.s. of (8.1.6') should be changed to:
\begin{multline}\label{eq:sigma_moment}
-\omega_{\alpha Z}\{\sigma_{\alpha Zlr}e_{slm}k_m\}=\sum_{\beta,\zeta}\frac{kT}{m_\alpha+m_\beta}\Bigg[\frac{7}{2}kT\mu_{\alpha \beta}\left(\frac{G^{(13)}_{\alpha z \beta \zeta}\pi_{\alpha zrs}}{m_\alpha^2p_{\alpha z}}+\frac{G^{(14)}_{\alpha z \beta \zeta}\pi_{\beta \zeta rs}}{m_\beta^2p_{\beta \zeta}}\right)\\
+\frac{G^{(15)}_{\alpha z \beta \zeta}\sigma_{\alpha zrs}}{p_{\alpha z}}+\frac{G^{(16)}_{\alpha z \beta \zeta}\sigma_{\beta \zeta rs}}{p_{\beta \zeta}}\Bigg]
\end{multline}
In Eq. (8.1.7):
\begin{align}
G^{(14)}_{\alpha Z \beta \zeta}=
-\frac{24}{35}\frac{m_\beta}{m_\alpha}\lambda_{\alpha Z \beta \zeta}\\
G^{(16)}_{\alpha Z \beta \zeta}=\frac{24}{7}\kappa_{\alpha\beta}\frac{m_\beta}{m_\alpha}\lambda_{\alpha Z \beta \zeta}
\end{align}
The coefficient $c^{(5)}_{\alpha}$ in Eq. (8.4.4) is:
\begin{align}\label{eq:c5}
c^{(5)}_{\alpha}=\frac{5}{2}\tau^{-1}_{\alpha}\tau_{\alpha\alpha}\frac{S^{(11)}_{\alpha}}{S^{(5)}_{\alpha}S^{(11)}_{\alpha}-7(S^{(9)}_{\alpha})^2}=\frac{5}{2}\tau^{-1}_{\alpha}\tau_{\alpha\alpha}\frac{S^{(11)}_{\alpha}}{D_\alpha}
\end{align}
The coefficient $c^{(6)}_{\alpha}$ in Eq. (8.4.4) is:
\begin{align}\label{eq:c6}
c^{(6)}_{\alpha}=\frac{S^{(8)}_{\alpha}S^{(9)}_{\alpha}-S^{(2)}_{\alpha}S^{(11)}_{\alpha}}{S^{(5)}_{\alpha}S^{(11)}_{\alpha}-7(S^{(9)}_{\alpha})^2}=\frac{S^{(8)}_{\alpha}S^{(9)}_{\alpha}-S^{(2)}_{\alpha}S^{(11)}_{\alpha}}{D_\alpha}
\end{align}
Add Boltzmann constant into (8.4.4):
\begin{align}\label{eq:heat_flux_expression}
\frac{\textbf h_{\alpha z}}{p_{\alpha z}}-\frac{\overline{\textbf h}_{\alpha}}{p_{\alpha}}=n_{\alpha}\tau_{\alpha}\tau^{-1}_{\alpha\alpha}c^{(5)}_{\alpha}\left(\frac{\overline{Z_\alpha^2}}{Z^2}k\nabla T_{\alpha z}-k\nabla T_{\alpha}\right)+c^{(6)}_{\alpha}(\textbf w_{\alpha z}-\overline{\textbf w}_{\alpha })
\end{align}
\begin{align}
\textrm{where}\ \tau^{-1}_{\alpha}=\sum_{\beta}\frac{\mu_{\alpha \beta}}{m_\alpha}\tau^{-1}_{\alpha \beta}\ \ \  \textrm{(power is corrected)}
\end{align}
\bibliographystyle{aip}
\bibliography{refs}
\end{document}